\documentclass[fleqn,usenatbib]{mnras}

\usepackage {comment}

\usepackage{newtxtext,newtxmath}
\usepackage[T1]{fontenc}
\usepackage{ae,aecompl,times}
\usepackage{graphicx}	% Including figure files
\usepackage{amsmath}	% Advanced maths commands
\usepackage{amssymb}	% Extra maths symbols
\usepackage{multirow}
\usepackage[normalem]{ulem}
\usepackage{xcolor}
\usepackage{bm}

\makeatletter{}\def\LCDM{\mbox{$\Lambda$CDM}}
\def\Mpch{\mbox{$h^{-1}$Mpc}}
\def\kMpch{\mbox{$h$Mpc$^{-1}$}}

\def\kpch{\mbox{$h^{-1}$kpc}}
\def\Gpch{\mbox{$h^{-1}$Gpc}}

\def\M200{\mbox{$M_{\rm 200 }$}}
\def\Msunh{\mbox{$h^{-1}M_\odot$ }}

\def\R200{\mbox{$R_{\rm 200 }$}}

\def\V200{\mbox{$V_{\rm 200 }$}}
\def\kms{\mbox{$\rm {km\,s}^{-1}$}}
\def\Mpcinv{\mbox{$\rm {Mpc}^{-1}$}}

\def\GLAM{\textsc{glam}\,\,}

\newcommand{\Ng}{\mbox{$N_{\rm g}$}}

\newcommand{\lsim}{\mbox{${\,\hbox{\hbox{$ < $}\kern -0.8em \lower 1.0ex\hbox{$\sim$}}\,}$}}
\newcommand{\gsim}{\mbox{${\,\hbox{\hbox{$ > $}\kern -0.8em \lower 1.0ex\hbox{$\sim$}}\,}$}}

\def\beqn{\vspace{2mm}
\begin{eqnarray}} 
\def\eeqn{\vspaceg{2mm} 
\end{eqnarray}}

\newcommand{\be}{\begin{equation}}
\newcommand{\ee}{\end{equation}}
\newcommand{\ba}{\begin{eqnarray}}
\newcommand{\ea}{\end{eqnarray}}
\newcommand{\brr}{\begin{array}}
 
\newcommand{\err}{\end{array}}
\newcommand{\bc}{\begin{center}}
\newcommand{\ec}{\end{center}}

\definecolor{mred}{rgb}{0.058, 0.588, 0.778}
 
\title[Clustering and Halo Abundances in EDE Cosmological Models]
{Clustering and Halo Abundances in Early Dark Energy Cosmological Models
}

\makeatletter{}\author[Klypin, Poulin, Prada, Primack et al.]
  {Anatoly~Klypin$^{1,2}$, Vivian~Poulin$^3$, Francisco~Prada$^4$,
    Joel~Primack$^5$, \newauthor
 Marc~Kamionkowski$^6$, Vladimir~Avila-Reese$^7$,  Aldo~Rodriguez-Puebla$^7$, Peter~Behroozi$^8$, \newauthor
 Doug Hellinger$^5$ and Tristan L.~Smith$^9$
   \vspace{0.2cm}\\ 
  $^1$Astronomy Department, New Mexico State University, Las Cruces, NM, USA\\
  $^2$Department of Astronomy, University of Virginia, Charlettesville, VA, USA\\
  $^3$Laboratoire Univers \& Particules de Montpellier, CNRS \& Universite de Montpellier, Montpelier, France\\
  $^4$ Instituto de Astrof\'{\i}sica de Andaluc\'{\i}a (CSIC), Glorieta de 
     la Astronom\'{\i}a, E-18080 Granada, Spain \\
  $^5$ Physics Department and SCIPP, University of California, Santa Cruz, CA 95064, USA\\
  $^6$Department of Physics and Astronomy, Johns Hopkins University, Baltimore, MD 21218, USA\\
  $^7$Instituto de Astronomıa, Universidad Nacional Autonoma de Mexico,  Ciudad de Mexico, Mexico\\
  $^8$Steward Observatory, University of Arizona, Tucson, AZ, USA %\\
}

\author[Klypin, Poulin, Prada, Primack et al.]{Anatoly~Klypin$^{1,2}$\thanks{E-mail: aklypin@nmsu.edu}, Vivian~Poulin$^3$, Francisco~Prada$^4$,
    Joel~Primack$^5$, \newauthor Marc~Kamionkowski$^6$, Vladimir~Avila-Reese$^7$, Aldo~Rodriguez-Puebla$^7$, Peter~Behroozi$^8$, \newauthor 
    Doug Hellinger$^5$ \& Tristan L.~Smith$^9$
   \vspace{0.2cm}\\ 
  $^1$Astronomy Department, New Mexico State University, Las Cruces, NM, USA\\
  $^2$Department of Astronomy, University of Virginia, Charlettesville, VA, USA\\
  $^3$Laboratoire Univers \& Particules de Montpellier, CNRS \& Universite de Montpellier, Montpelier, France\\
  $^4$ Instituto de Astrof\'{\i}sica de Andaluc\'{\i}a (CSIC), Glorieta de 
     la Astronom\'{\i}a, E-18080 Granada, Spain \\
  $^5$ Physics Department and SCIPP, University of California, Santa Cruz, CA 95064, USA\\
  $^6$Department of Physics and Astronomy, Johns Hopkins University, Baltimore, MD 21218, USA\\
  $^7$Instituto de Astronomıa, Universidad Nacional Autonoma de Mexico,  Ciudad de Mexico, Mexico\\
  $^8$Steward Observatory, University of Arizona, Tucson, AZ, USA \\
  $^9$Department of Physics and Astronomy, Swarthmore College, 500 College Ave., Swarthmore, PA 19081, USA %\\
}

\begin{document}

\maketitle
\label{firstpage}
\begin{abstract}
%Abstract revised by Joel
\LCDM\ cosmological models with Early Dark Energy (EDE) have been proposed to resolve tensions between the Hubble constant $H_0 = 100 h$ km s$^{-1}$ Mpc$^{-1}$ measured locally, giving $h \approx 0.73$, and $H_0$ deduced from Planck cosmic microwave background (CMB) and other early universe measurements plus \LCDM, giving $h \approx 0.67$.  EDE models do this by adding a scalar field  that temporarily adds dark energy equal to about 10\% of the cosmological energy density at the end of the radiation-dominated era at redshift $z \sim 3500$.  Here we compare linear and nonlinear predictions of a Planck-normalized \LCDM\ model including EDE giving $h = 0.728$ with those of standard Planck-normalized \LCDM\ with $h =0.678$. We find that nonlinear evolution reduces the differences between power spectra of fluctuations at low redshifts.  As a result, at $z=0$ the halo mass functions on galactic scales are nearly the same, with differences only 1-2\%. However, the differences dramatically increase at high redshifts. The EDE model predicts 50\% more massive clusters at $z=1$ and twice more galaxy-mass halos at $z=4$. Even greater increases in abundances of galaxy-mass halos at higher redshifts may make it easier to reionize the universe with EDE.  Predicted galaxy abundances and clustering will soon be tested by JWST observations.
Positions of baryonic acoustic oscillations (BAOs) and correlation functions differ by about 2\% between the models -- an effect that is not washed out by nonlinearities. 
Both standard \LCDM\ and the EDE model studied here agree well with presently available acoustic-scale observations, but DESI and Euclid measurements will provide stringent new tests.
\end{abstract}

\begin{keywords}
cosmology: Large scale structure - dark matter - galaxies: halos - methods: numerical
\end{keywords}

\makeatletter{}\section{Introduction}
\label{sec:intro}

\begin{figure} 
\centering
\includegraphics[scale = 0.5, width=0.49\textwidth]
{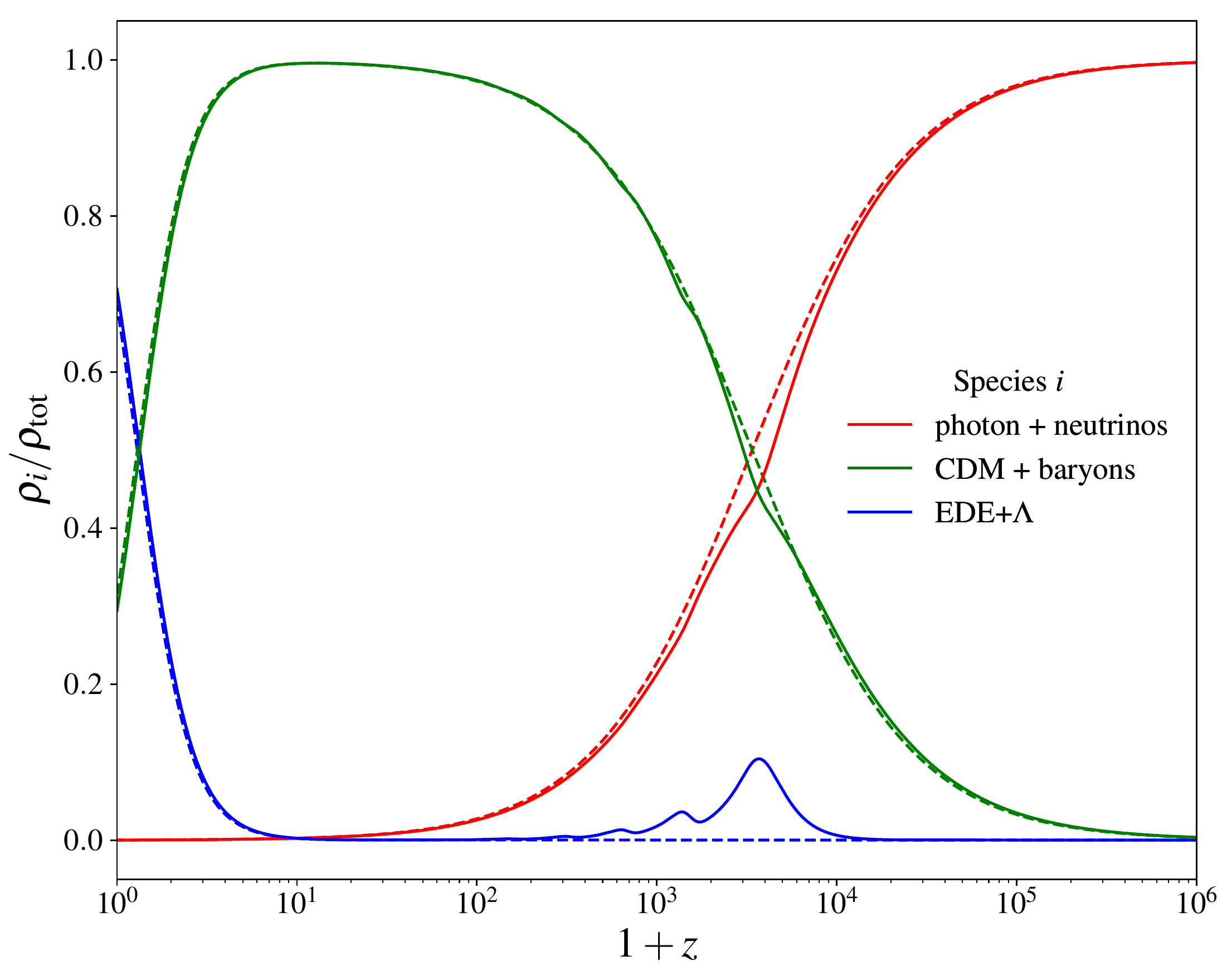}
\caption{Densities of different components at different redshifts for
  EDE (full curves) and the standard \LCDM\ model (dashed
  curves). Oscillating early dark energy density (blue curve) peaks at
  $z\sim 3500$ when it contributes $\sim 10\%$ to the total
  density. Its contribution quickly decreases after that.  }
\label{fig:densities}
\end{figure}

Combined late-universe measurements give the value of the Hubble 
constant %$H_0 = 73.3\pm 0.8$ (in the usual units of km/s/Mpc)
$h = 0.733\pm0.008$
according to a recent review of \citet{Verde2019}. This value of the
expansion rate is in as much as $6\sigma$ conflict with the value
%$H_0 = 67.4\pm 0.5$ 
$h=0.674\pm0.005$ from the Planck measurements of the cosmic
background radiation (CMB) temperature and polarization and other
early-universe observations extrapolated to the present epoch using
standard $\Lambda$CDM \citep{Planck2018}. This discrepancy is unlikely
to be a statistical fluke, and it is not easily attributable to any
systematic errors
\citep[e.g.,][]{freedman2017cosmology,Riess2019,Aylor2019}. Instead,
it may be telling us that there is a missing ingredient in standard
\LCDM.  Of the many potential explanations that have been proposed, a
brief episode of early dark energy (EDE) around the time of matter
dominance followed by \LCDM\ evolution
\citep{Poulin2019,Knox2020,SmithEDE2020,Argawal2019,Lin2019} has received perhaps the most attention.
For the model we consider here, \citet{Poulin2019} and \citet[] [SPA20]{SmithEDE2020} have shown that their fluctuating scalar field EDE model can fit all the
CMB data as well as the usual standard 6-parameter \LCDM\ does, and
also give $H_0$ in agreement with the recent local-universe
measurements. As Figure~\ref{fig:densities} shows, in this model the early dark
energy contributes a maximum of only about 10\% to the total cosmic
density at redshifts $z\sim 3500$, at the end of the era of radiation domination and the beginning of matter domination.

The resulting best-fit cosmic parameters (see
Table~\ref{table:cosmology}) are interestingly different from those of
standard \LCDM. In particular, both the primordial power spectrum
amplitude $A_s$ and $\sigma_8$, measuring the linear amplitude today at
$8\Mpch$, are larger than for the latest Planck analysis with standard
\LCDM. Also, $n_s$, the slope of the power primordial power spectrum
is larger than for standard \LCDM. And with the higher $H_0$, the present age of the universe is 13.0 Gyr rather than 13.8 Gyr.  Such modifications of the cosmological parameters are also produced in other recent
papers on EDE \citep{Argawal2019,Lin2019}. 

Particle theory provides many scalar fields that could have nonzero
potential energy temporarily preserved by Hubble friction, 
leading to temporary episodes of effective dark
energy \citep[e.g.,][]{Dodelson2000,Griest2002,Kamionkowski2014}. It has long been known that dark
energy contributions at early cosmic times can imply modifications of
CMB, big-bang nucleosynthesis, and large-scale structure formation
\citep{Doran2001,Muller2004,Bartelmann2006}.

Only recently has resolving the Hubble tension become a motivation for
EDE \citep{Karwal2016,Poulin2019}. The challenge
lies in finding ways in which the Hubble parameter inferred from the
CMB can be made larger without introducing new tensions with the
detailed CMB peak structure and/or other well established cosmological
constraints. In particular, all solutions are constrained by the
remarkable precision (roughly one part in $10^4$) with which the angular
scale $\theta_a$ of the acoustic peaks in the CMB power spectrum is
fixed. Roughly speaking, this angular scale is set by $\theta_a\propto r_s/D_A$,
where $r_s$ is the comoving sound horizon at the surface of last scatter
and $D_A$ is the comoving distance to the surface of last scatter. 

There are two possibilities to keep $\theta_a$ fixed: keep $D_A$ fixed
by compensating the increase of energy today ($H_0$ higher means
higher energy density today) by decreasing the energy density at
earlier times through a change to the late-time expansion history, or
decreasing $r_s$ by the same amount as $D_A$ through a change to the
early-time physics. However, modifications to the late-time expansion
history are constrained by measurements of baryonic acoustic
oscillations and luminosity distance to supernovae, and early-time solutions are constrained by the detailed
structure of the higher acoustic peaks in the CMB power spectra
\citep{Bernal2016}. Even so, \citet{Poulin2018} and subsequent studies
\citep{Poulin2019,Argawal2019,SmithEDE2020,Lin2019} were able to find regions of
the parameter space of EDE models that provide a good fit to the
data. Still, more work must be done — both in terms of theory and new
measurements — to assess the nature of viable EDE models.

%\Vivian{VP: I believe we should clearly state the questions we wish to answer in this paper: Something like "In this paper, we compare for the first time the prediction for LSS observables between LCDM and EDE. Through a suite of non linear simulations, we compute the halo mass function and the baryonic acoustic oscillation (and correlation function) at low redshifts. We find significant differences that will allow future experiments such as JWST to disentangle between both cosmologies."} -- Joel: Inserted 3 paragraphs lower.

We have chosen to focus on the \citet{SmithEDE2020} version of EDE
because it was engineered to fit the details of the high-$l$ CMB
polarization data, and because it represents the best fit to the local
$H_0$ measurements and the largest deviation of the cosmological
parameters from standard \LCDM, which should lead to the clearest
differences in testable predictions. These new cosmological models
will make specific predictions for galaxy mass and luminosity
functions and galaxy clustering. Given that these phenomena arise from
nonlinear evolution of primordial perturbations and involve gas
dynamics, the power of numerical simulations is essential.  Of course,
it is possible that the result of such observational tests of EDE will
be to eliminate this class of cosmological models. But if not, EDE
potentially tells us about a phenomenon that contributes to early
cosmic evolution, and about another scalar field important in the
early universe besides the putative inflaton responsible for the
cosmic inflation that set the stage for the Big Bang.

There were some earlier efforts to study effects of nonlinear evolution in
models called early dark energy 
\citep{Bartelmann2006,Grossi2009,Fontanot2012,Francis2009}. However,
models for the dark energy used in those papers are very different as
compared with those discussed in this paper. As a matter of fact,
there is little in common -- with the exception of the name EDE --
between those models and the model we consider here.  The equation of state $w$ of dark
energy $P=w\rho c^2$ in those papers is $w=-1$ only at $z=0$ and has
significant deviations from $w=-1$ at low redshifts. For example,
models used by \citet{Grossi2009} and \citet{Fontanot2012} had
$w=-0.7$ at $z=1$ and $w=-0.4$ at $z=5$. This should be compared with
$w=-1$ at $z \lesssim 1000$ in our EDE model.
%$w\to 1/2$ for all redshifts $z\lesssim1000$ and $w\to -1$ at $z\gg 300$ 
%See June 9 email from Anatoly Klypin explaining that this is about the total w, not just the contribution from EDE

In the sense of dynamics of growth of fluctuations in the matter-dominated era in our EDE model,
we are dealing with a vanilla \LCDM\ model with the  only modification being
the spectrum of fluctuations. Even the spectrum of fluctuations is not
much different: a 2\% change in $\sigma_8$ and 0.02 difference in
the slope of the spectrum. With these small deviations, one might imagine that the final non-linear statistics (such as power, correlation functions, halo mass functions) would be very similar.  But instead we find very significant differences, especially at redshifts $z > 1$.

The $S_8$ tension is the conflict between weak lensing and other local observations that imply a relatively low value of $S_8 \equiv \sigma_8 \sqrt{\Omega_m/0.3}$ and the higher value of $S_8$ of both the Planck-normalized \LCDM\ and the EDE model considered here \citep{SmithEDE2020}.  
Our EDE model has $\sigma_8 = 0.836$, larger than $\sigma_8 = 0.820$ of our fiducial Planck 2013 MultiDark model or the Planck 2018 value $\sigma_8 = 0.811\pm0.006$. But what is determined by CMB observations is $\Omega_m h^2$, and the higher value of $H_0$ with EDE means that the resulting $S_8 = 0.830$ is identical to that from Planck 2018 \citep[Combined value, Table 1 of][]{Planck2018}. 

The latest weak lensing measurements of $S_8$ are the Dark Energy Survey year 1 (DES-Y1) cosmic shear results $S_8 = 0.782^{+0.027}_{-0.027}$ \citep{Troxel2018}; the Hyper Suprime-Cam Year 1 (HSC-Y1) cosmic shear power spectra, giving $S_8 = 0.800^{+0.029}_{-0.028}$ \citep{Hikage2019}; and the HSC-Y1 cosmic shear two-point correlation functions, giving $S_8 = 0.804^{+0.032}_{-0.029}$ \citep{Hamama2020}. These measurements are all in less than $2\sigma$ disagreement with $S_8 = 0.830$ from Planck-normalized \LCDM\ and our EDE model.

\citet{Hill2020} claims that the EDE model considered here, and other EDE models,  are in serious tension with large scale structure measurements.  They cite the DES-Y1 result $S_8 = 0.773^{+0.026}_{-0.020}$, obtained by combining weak lensing with galaxy clustering \citep{Abbott2019}, which disagrees by $2.3\sigma$ with $S_8 = 0.830$. However, \citet{Abbott2019} allowed the total neutrino mass free to vary, which leads to a somewhat lower DES-inferred $S_8$ than that, $S_8=0.792\pm0.024$, which arises if $\sum m_\nu=0.06$ eV is fixed, as the Planck team \citep{Planck2018} and we have done. Similarly, the shear only result was analyzed by the SPTPol collaboration with the same convention as ours; they obtained $S_8=0.79^{+0.4}_{-0.029}$ \citep{Bianchini:2019vxp}, to be compared with $S_8 = 0.782\pm0.027$ once the sum of neutrino masses is left free to vary \citep{Troxel2018}. While there is indeed some $S_8$ tension between the DES-Y1 measurements and the prediction of our EDE model, it remains true that the addition of a brief period of early dark energy resolves the \LCDM\ Hubble tension and fits the Planck 2018 CMB observations without exacerbating the $S_8$ tension. This is confirmed from table 7 of \citet{Hill2020}, where one can read off that the joint DES-Y1 $\chi^2$ goes from $506.4$ within $\Lambda$CDM to $507.7$ in the EDE cosmology, a marginal degradation given that the joint DES-Y1 data have $457$ data points \citep{Abbott2019}. This allows us to conclude that the DES-Y1 result does not exclude the presence of EDE.  Further measurements by DES, HSC, and other programs will be important tests for cosmological models as they improve the precision of measurements of $S_8$ and other cosmological parameters.

In this paper, we compare for the first time the predictions for large scale structure observables between standard \LCDM\ and EDE. Through a suite of non-linear simulations, we compute the halo mass function and the baryonic acoustic oscillations (and correlation functions) at various redshifts. We find significant differences that will allow future observations such as those from eROSITA, JWST, DESI, and Euclid to critically test such cosmologies.

\begin{figure}
  \centering
\includegraphics[width=0.52\textwidth]
{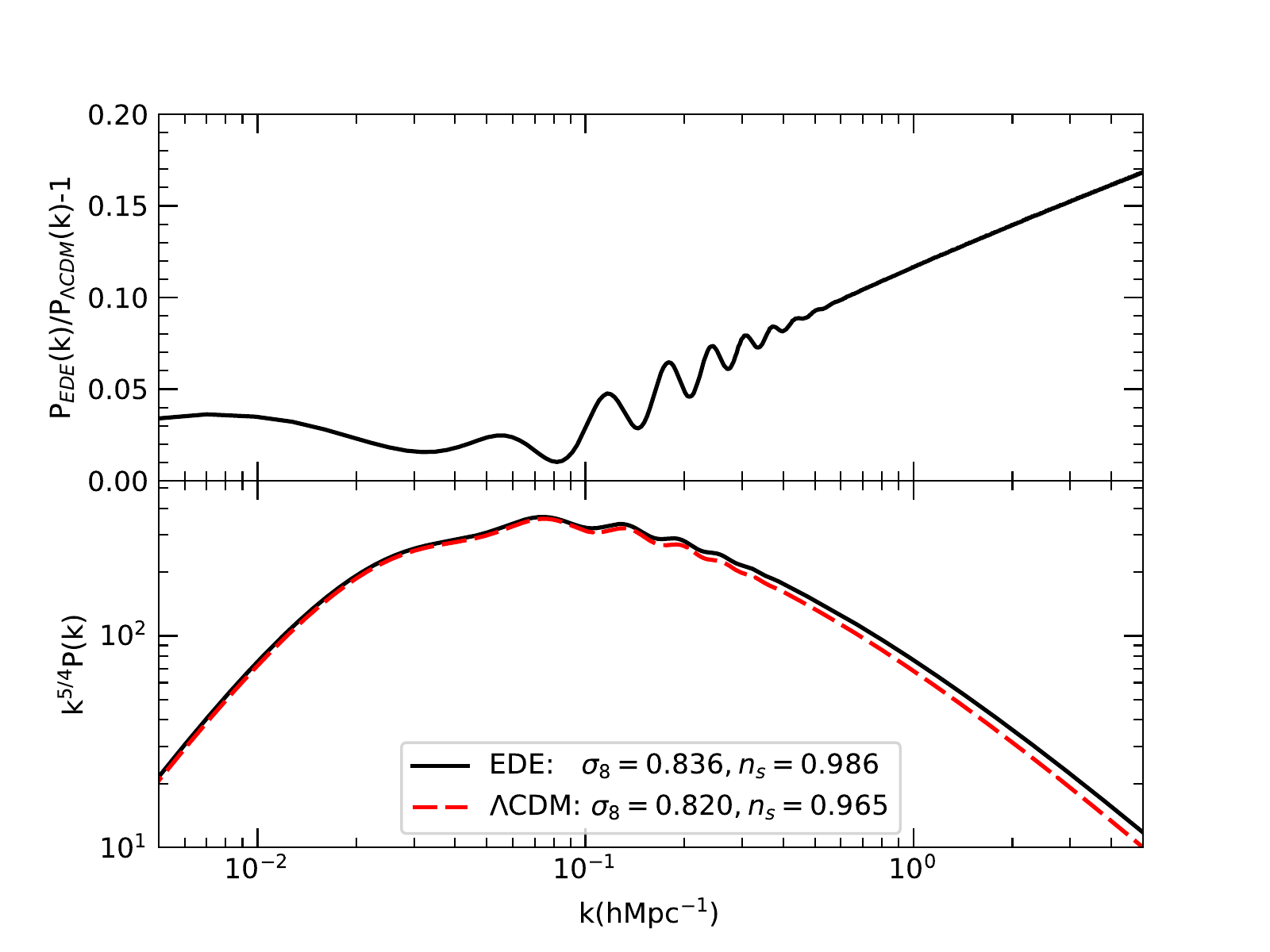}
\caption{{\it Bottom panel:} Linear power spectrum of dark matter
  fluctuations at $z=0$ scaled with factor $k^{5/4}$ to reduce the
  dynamical range and to make the domain of BAOs
  $k= (0.07-0.3)h{\rm Mpc}^{-1}$ more visible. {\it Top panel:} The
  ratio of power spectra in our EDE model to that of the standard \LCDM\
  model.  The amplitude of fluctuations in our EDE model is always
  larger than in \LCDM\ though the differences at long wavelengths
  $\lsim 0.1h{\rm Mpc}^{-1}$ are only (2--3)\%. The differences
  increase at large $k$ and become substantial ($\sim 20\%$) on
  galactic scales $k\gsim 5h{\rm Mpc}^{-1}$. }
\label{fig:power}
\end{figure}

We use extensive $N$-body simulations to study the effects of non-linear
evolution. As a benchmark, we employ a \LCDM\ model with the parameters
and spectrum of the MultiDark-Planck simulations
\citep{Klypin2016,Rodriguez-Puebla2016}. Table~\ref{table:cosmology} lists those parameters
and Figure~\ref{fig:power} compares linear power spectra.  
%There are reasons to use the MultiDark model.  It 
MultiDark-Planck is a well studied \LCDM\ model based on the 2013 Planck cosmological
parameters \citep{Planck2013}
that has been used in many publications. Sophisticated analyses of galaxy
statistics applied to different MultiDark-Planck numerical simulations show that the model reproduces the observed clustering of
galaxies in samples such as SDSS and BOSS
\citep[e.g.,][]{Guo2015,Sergio2016,BOSSmocks2016}. Analyses of this kind --
matching selection functions, boundaries of observational sample,
light cones, and stellar luminosity functions -- are difficult to
implement and require high-resolution simulations. We plan to do such simulations in the future for the EDE model considered here,
but for now we are interested in learning what differences to expect
and what statistics should be promising to distinguish between standard \LCDM\ models compared with with EDE ones.

In \S2 we describe the cosmological simulations used in this paper, and in \S3 we present and discuss the resulting power spectra. In \S4 we compare the baryon acoustic oscillations and corresponding correlation functions between \LCDM\ and the EDE model. In \S5 we discuss the changes in halo abundances in EDE out to redshift $z =4$, and explain the origin of these changes. In \S6 we discuss halo abundance and clustering at even higher redshifts, including implications for reionization of the universe.  \S7 is a summary and discussion of our results.

\makeatletter{}

\makeatletter{}

\makeatletter{}\begin{table}
\begin{center}
\caption{Parameters of cosmological models.}
\begin{tabular}{ l | c | c | c}
\hline\hline  
Parameter & EDE &  \LCDM      & \LCDM \\
          &  SPA20   &  MultiDark-Planck13 & CMB-Planck18  \\
\hline  
$\Omega_{\rm m}$             & 0.293     & 0.307    & 0.315$\pm 0.007$ \\ 
$\Omega_{\rm cold}h^2$ & 0.132     & 0.119    & 0.120 $\pm 0.001$ \\
$\Omega_{\rm bar}h^2$  & 0.0225    & 0.0221   & 0.0224$\pm 0.0001$ \\
$H_{\rm 0}$ [\kms \Mpcinv]                   & 72.81    & 67.77   & 67.36$\pm 0.54$ \\
$n_{\rm s}$                  & 0.986     & 0.965    & 0.965$\pm 0.004$ \\ 
$\sigma_8$             & 0.836     & 0.820    & 0.811$\pm 0.006$ \\ 
Age [Gyr]              & 13.032    & 13.825   & 13.797$\pm 0.023$ \\
$z_{\rm drag}$         &   1061.28  &   1059.09  &  1059.94$\pm 0.30$  \\
$r_{\rm drag}$ [Mpc]            & 140.1     & 147.8    & 147.1$\pm 0.3$ \\
\tabularnewline
\hline
\end{tabular}
\label{table:cosmology}
\vspace{-5mm}
\end{center}
\end{table}

\makeatletter{}\begin{table*}
\begin{center}
\caption{Numerical and cosmological parameters of different simulations.
  The columns give the simulation identifier, cosmology,
  the size of the simulated box in $h^{-1}\,{\rm Mpc}$,
  the number of particles, 
  the mass per simulation particle $m_p$ in units of $h^{-1}\,M_\odot$, the mesh size $\Ng^3$,
  the  gravitational softening length $\epsilon$ in units of $h^{-1}\,{\rm Mpc}$, the number of time-steps $N_{\rm step}$, initial redshift, and
  the number of realizations $N_r$.  Additional smaller-scale simulations are discussed in \S6.}
\begin{tabular}{ l | c | r | r |   r | r |  r| r |r |r }
\hline\hline  
Simulation & Cosmology & Box\phantom{1} & particles  & $m_p$\phantom{mmm}   & $\Ng^3$  & $\epsilon$ \phantom{m}  & $N_{\rm s}$ & $z_{\rm init}$  & $N_r$ 
\tabularnewline
  \hline 
EDE$_{0.5}$   & EDE       & 500$^3$    & 2000$^3$ & $1.3\times 10^9$   & 7000$^3$ & 0.071 & 253 & 150   & 5 \\
EDE$_{1}$  & EDE       & 1000$^3$   & 2000$^3$ & $1.0\times 10^{10}$ & 7000$^3$ & 0.143 & 136 & 100   & 16 \\
EDE$_{2A}$  & EDE       & 2000$^3$   & 2000$^3$ & $8.3\times 10^{10}$ & 7000$^3$ & 0.285 & 130 & 150   & 6 \\
EDE$_{2B}$  & EDE       & 2000$^3$   & 2000$^3$ & $8.3\times 10^{10}$ & 4000$^3$ & 0.500 & 130 & 150   & 210 \\
$\Lambda$CDM$_{0.5}$   & MultiDark       & 500$^3$    & 2000$^3$ & $1.3\times 10^9$   & 7000$^3$ & 0.071 & 253 & 150   & 5 \\
$\Lambda$CDM$_{1}$  & MultiDark       & 1000$^3$   & 2000$^3$ & $1.1\times 10^{10}$ & 7000$^3$ & 0.143 & 136 & 100   & 30 \\
$\Lambda$CDM$_{2A}$  & MultiDark       & 2000$^3$   & 2000$^3$ & $8.3\times 10^{10}$ & 7000$^3$ & 0.285 & 130 & 100   & 15 \\
$\Lambda$CDM$_{2B}$  & MultiDark       & 2000$^3$   & 2000$^3$ & $8.3\times 10^{10}$ & 4000$^3$ & 0.500 & 130 & 150   & 210 \\
%\tabularnewline

\tabularnewline
\hline
\end{tabular}
\label{table:simtable}
\vspace{-5mm}
\end{center}
\end{table*}

\section{Simulations}
\label{sec:sim}

Most of the results presented in this paper are based on new cosmological
$N$-body simulations.  The simulations were carried out with the
parallel Particle-Mesh code \GLAM~\citep{GLAM}. Because the \GLAM\
code is very fast, we have done many realizations of the simulations
with the same cosmological and numerical parameters that only differ
by the initial random seed. A large number of realizations is quite important
because the differences between EDE and \LCDM\ models are not very
large. This is especially true on long wavelengths $ k\lsim 0.1\kMpch$ where
the difference in the power spectra is just $\sim 2$\%.  So, one needs
many realizations to reduce the cosmic variance and see the real
differences.

  All the \GLAM~simulations were started at initial redshift
$z_{\rm init}=100$ or $z_{\rm init}=150$ using the Zeldovich
approximation.
Table~\ref{table:simtable} presents the numerical parameters of
our simulation suite: box-size, number of particles, particle mass 
$m_p$, number of mesh points $N_g^3 $, cell-size of the
density/force mesh $\epsilon$, the number of time-steps
$N_{\rm step}$, and the number of realizations $N_r$.

The \GLAM\ code is very fast as compared with high-resolution codes such
as \textsc{gadget} \citep{GADGET} or \textsc{ART} \citep{ART}. For
example, our most expensive simulations EDE$_{0.5}$ and
$\Lambda$CDM$_{0.5}$ used just $\sim 2500$~cpu-hours on a dual Intel
Platinum 8280M computational node, which is just 2 days of wall-clock
time. The limiting factor of \GLAM\ simulations is the force
resolution $\epsilon$. It is defined by the cell size - the ratio of the box size
$L$ to the mesh size $N_g$: $\epsilon = L/N_g$. So, the
larger the mesh size $N_g$, the better is the resolution.
\citet{GLAM} give detailed analysis of convergence and accuracies of
the \GLAM\, code. Just as with any Particle-Mesh code, the resolution
is defined by the available memory: the larger the memory, the better
the resolution. We use computational nodes each with 1.5Tb RAM and
two Intel Platinum 8280M processors with combined 56 cores.

We use a spherical overdensity (SO) halo finder, which is a stripped
down variant of the Bound Density Maxima (BDM) halo finder \citep{Bolshoi,Knebe2011}.  Limited force
resolution does not allow subhalos to survive in virialized
halos. This is why we study only distinct halos (those that are not subhalos) in the present paper.  

\makeatletter{}\begin{figure}
  \centering

\includegraphics[width=0.52\textwidth]
{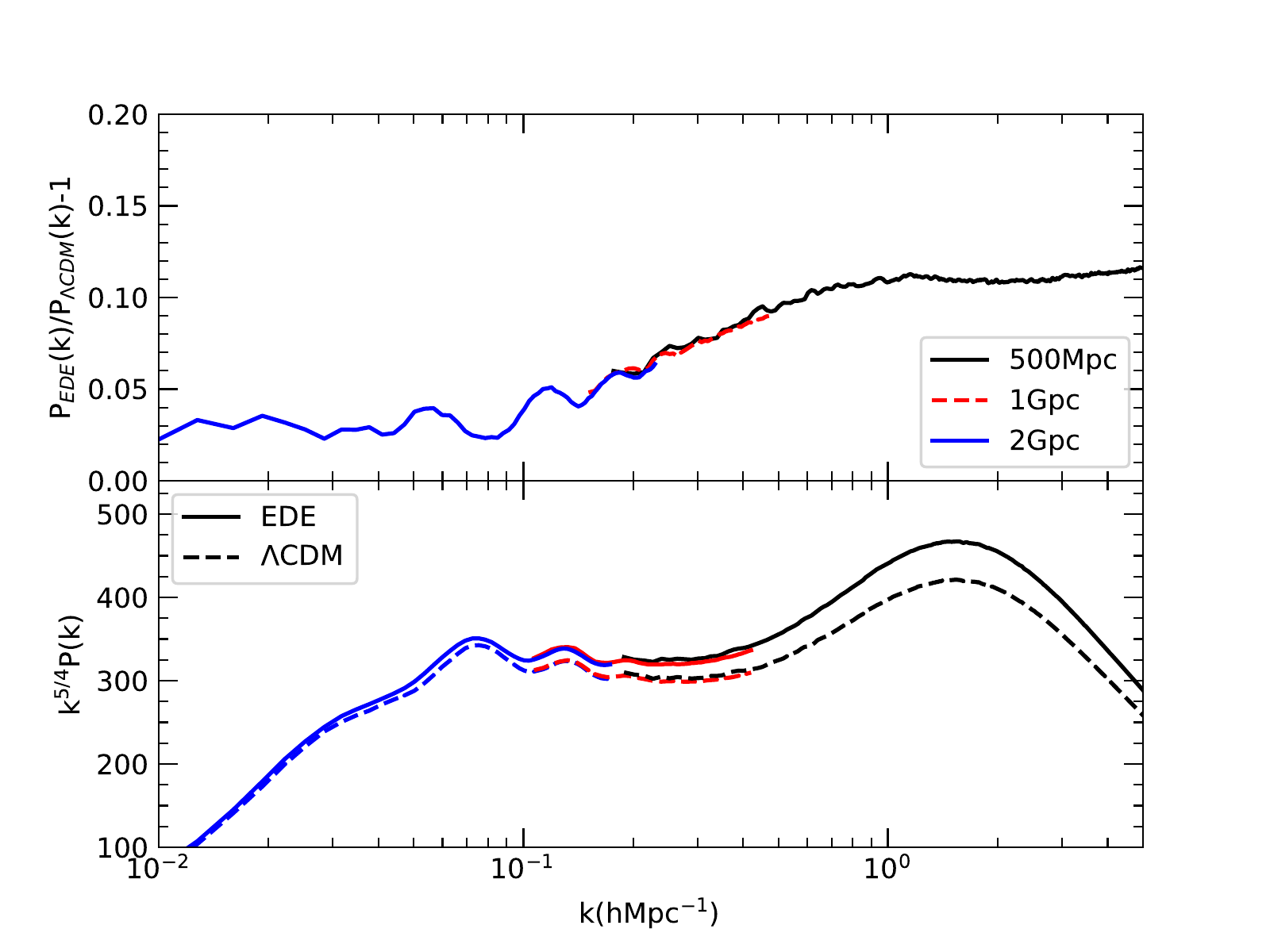}
\caption{Similar to Figure~\ref{fig:power} but for nonlinear evolution
  at $z=0$. Results from different box sizes and resolutions nicely
  match in overlapping regions. Nonlinear evolution dramatically
  changes the shape of the power spectrum at small scales. The BAO
  peaks are slightly damped, broadened and shifted. To some degree the nonlinear
  effects reduce  the differences between the models,
  but they do not wipe them out. }
\label{fig:powerNL}
\end{figure}

\makeatletter{}\begin{figure*}
  \centering
\includegraphics[width=0.82\textwidth]
{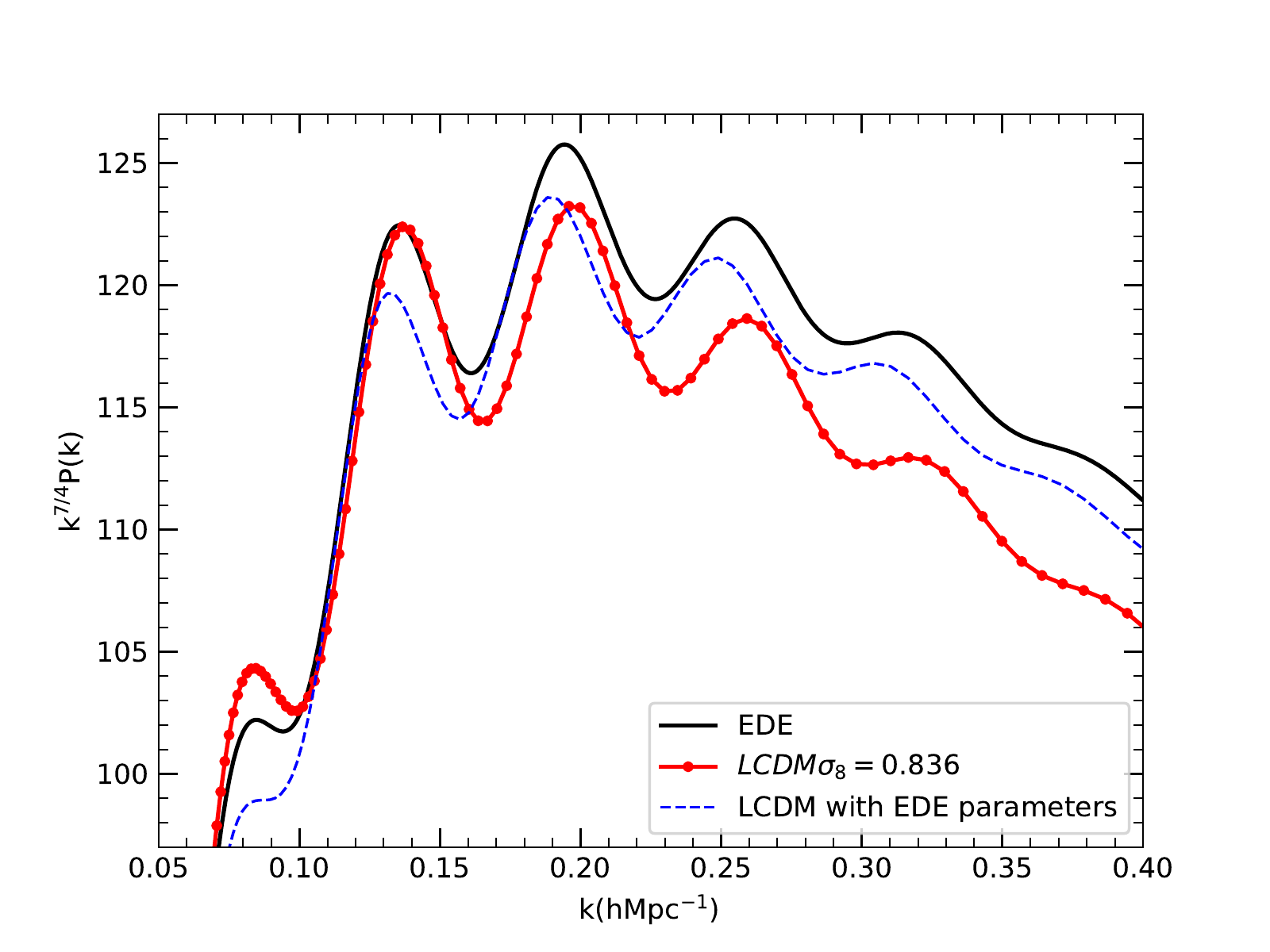}
\caption{Linear power spectra, scaled with a factor k$^{7/4}$, for the EDE (solid line) and \LCDM\ (dot-solid line) models in the BAO domain.
A third \LCDM\ model with the same cosmological parameters as EDE but without the early dark matter component 
is also shown (dashed line). All models were normalized to have the same $\sigma_8=0.836$ to appreciate
clearly the overall shape and acoustic oscillation features differences.
}
\label{fig:PkBAO}
\end{figure*}

\section{Power Spectra}
\label{sec:power}
Figure~\ref{fig:power} shows the $z=0$ linear power spectra of fluctuations
in the EDE and \LCDM\ models.  Differences between power spectra of
fluctuations are relatively small. On long wavelengths ($k\lsim 0.1\kMpch$)
the differences are mostly explained by the
normalizations: $\left[\sigma_8(EDE)/\sigma_8(\LCDM)\right]^2=1.039$.  The
differences increase on small scales and become substantial. For
example, at $k=5\kMpch$ the amplitude of fluctuations in the EDE model
is 17\% bigger than in the \LCDM\ model. 

The reason for this increase comes from the differences in the slope
$n_s$ of the primordial power spectra. At first sight the difference
of 0.02 in the slope seems to be small. However, it results in large
differences in amplitude when one compares waves that differ
dramatically in wavelength: 15\% for waves that differ by a factor of 1000 in
wavelength. A more subtle effect 
is related to the halo
mass function, which depends not only on the amplitude of fluctuations
but also on the slope of the power spectrum.

The domain of BAOs ($k=0.07-0.3\kMpch$) is also different in the
models. At first glance, the wiggles that are clearly seen in the top
panel of Figure~\ref{fig:power} are the familiar BAOs. They are not,
though they are related to BAOs. If the positions of the BAO peaks were the
same, there would not have been wiggles in the ratio of the power spectra.
Without the early dark energy component the position of BAO peaks is
mostly defined by $\Omega_{\rm bar}/\Omega_{\rm m}$ and
$\Omega_{\rm m}h^2$ \citep{Eisenstein1998}.  There is an additional
effect in EDE models due to the fact that the early dark energy
changes the dynamics of acoustic waves before the recombination.  So,
the very presence of the wiggles tells us that BAO peaks happen at
different wavenumbers: in the EDE models the BAOs are shifted to
slightly smaller wavenumbers. 

Nonlinear evolution modifies the power
spectra. Figure~\ref{fig:powerNL} shows results of our simulations at
redshift $z=0$. Results from different box sizes and resolutions
nicely match each other in overlapping regions. As the result, we stack
together different simulations and extend the range of resolved
scales.

As clearly seen in Figure~\ref{fig:powerNL} the nonlinear evolution
dramatically changes the shape of the power spectrum: at
$k\gsim 0.5\kMpch$ the fluctuations are much larger as compared with
the linear regime. The bump at $k\sim 1.5\kMpch$ corresponds to mass
$M=(4\pi/3)\Omega_m\rho_{\rm cr}(\lambda/2)^3\approx 
10^{13}\Msunh$
-- scale of large galaxies like our Milky Way. So, the bump is a
manifestation of collapsing dark matter halos.\footnote{There is no
  real peak in the power spectrum at those wave-numbers. The peak at
  $k\approx 1.5\kMpch$ in Figure~\ref{fig:powerNL} is due to the fact
  that we scale the power spectrum by factor $k^{5/4}$. However, there
  is a significant change in the slope of the power spectrum from
  $P(k)\propto k^{-2.5}$ in the linear regime to much flatter
  $P(k)\propto k^{-1.25}$.}

To some degree the nonlinear effects
reduce the differences between the models at strongly non-linear
regime $k\gsim 1\kMpch$. Here the ratios of the power spectra are
nearly constant 10\% -- a marked deviation from the linear
spectra shown in Figure~\ref{fig:power}. This nearly constant ratio of non-linear spectra produces
small and hardly detectable differences in the abundance of halos at
$z=0$.  Note that at larger redshifts the differences are larger than
at $z=0$ because the nonlinearities are smaller.

The power spectra in the domain of BAOs are also affected by nonlinearites,
but in a more subtle way.  The BAO peaks are slightly damped,
broadened and shifted: effects that are well understood and well
studied \citep[e.g.,][]{Eisenstein2007,AnguloBAO,Prada2016}, see Section~\ref{sec:BAO} for a detailed study. At even larger 
scales $k\lsim 0.05\kMpch$ the fluctuations are still in the nearly linear regime. 

The fact that nonlinear evolution reduces differences between EDE and
\LCDM\ models is a welcome feature. We know that at low redshifts
$z\lsim 0.5$ the \LCDM\ model reproduces the observed clustering of
galaxies in samples such as SDSS and BOSS
\citep{Guo2015,Sergio2016,BOSSmocks2016}. So, too large deviations
from \LCDM\ may point to problems. Nevertheless, though relatively
small, the deviations still exist and potentially can be
detected. The fact that nonlinear evolution reduces the differences
implies that one also expects larger differences at higher redshifts.
Indeed, this is what we find from analysis of halo abundances discussed below.

\section{Baryonic Acoustic Oscillations}
\label{sec:BAO}

\makeatletter{}\begin{figure}
  \centering
\includegraphics[width=0.49\textwidth]
{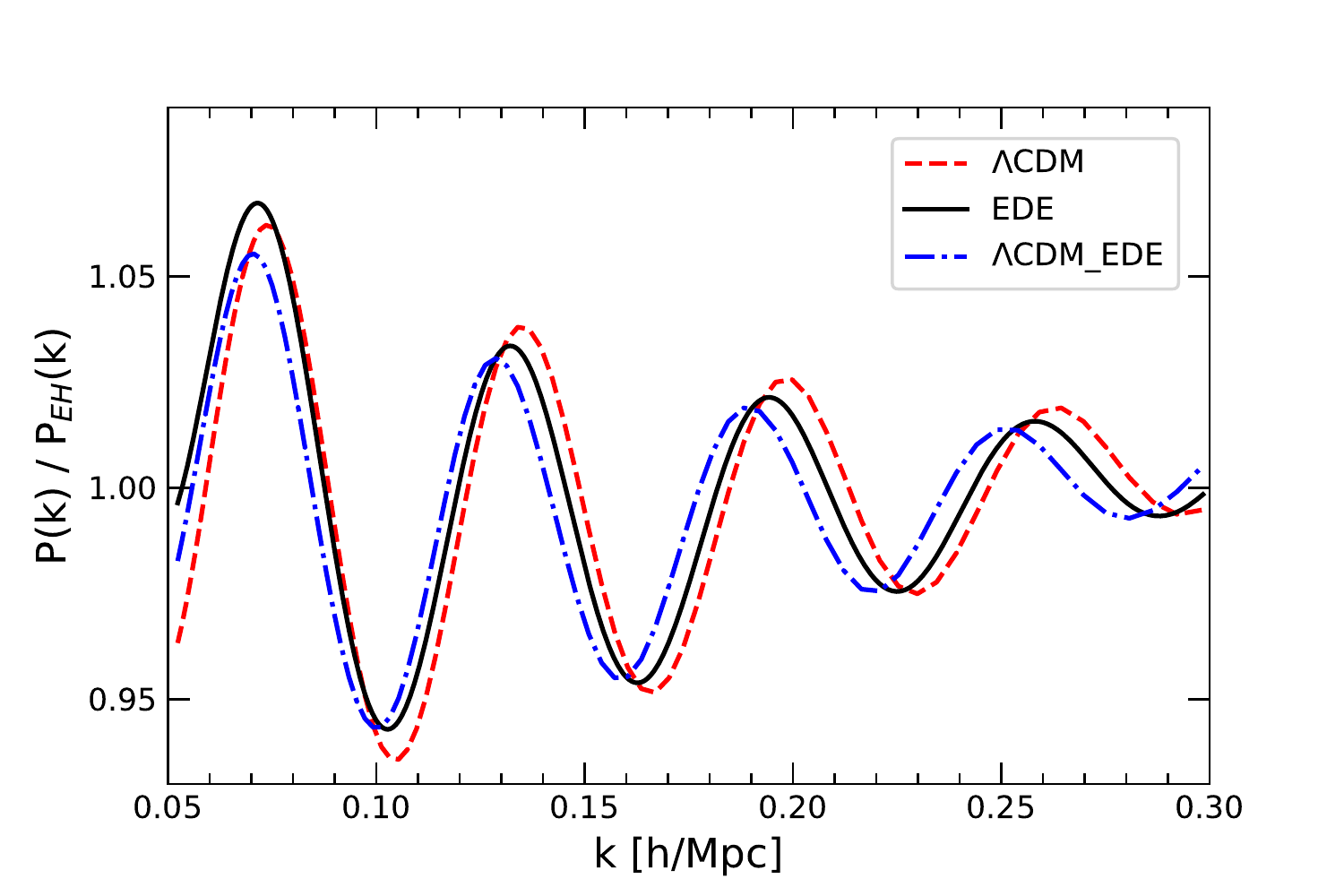}
\caption{BAO wiggles in the linear power spectrum for the three
  cosmological models: EDE (solid line), \LCDM\ (dashed line), and \LCDM\_EDE (dot-solid line) with the same
  cosmological parameters as EDE. The plot shows the deviations of the power
  spectra from that without baryonic oscillations \citep{Eisenstein1998}. As compared with \LCDM\ the BAO peaks in EDE are systematically 
  shifted by $1.8\%$ to smaller wavenumbers, and in the case of \LCDM\_EDE by $4.4\%$ to smaller wavenumbers.}
\label{fig:BAOwiggles}
\end{figure}

Figure~\ref{fig:PkBAO} displays the linear power spectra for the EDE (solid line) and two \LCDM\ models
%, with cosmological parameters listed in  Table~\ref{table:cosmology}, 
in the domain of the BAO features. In order to appreciate more clearly their 
overall P(k) shapes and BAO differences, the two \LCDM\ models have been normalized to have the same $\sigma_8=0.836$ as that of EDE.  One \LCDM\ model is otherwise the MultiDark-Planck one (dot-solid line).
The other \LCDM\ model (named \LCDM\_EDE, dashed line) has the same cosmological 
parameters as EDE but without the effects of the early dark energy component.

As compared with \LCDM\, the BAO peaks in the EDE model are systematically shifted to smaller
wavenumbers. This reflects the fact that the acoustic sound horizon has a larger value due to the faster expansion before the epoch 
of recombination. On the other hand, the sound horizon $r_{\rm d}$ in the EDE cosmology is smaller as compared to 
the \LCDM\_EDE model ($r_d=143.92$ Mpc) despite both cosmologies having the same cosmological parameters, and hence the BAO peaks in the latter are shifted towards larger scales. 
%This contradicts the fact that 
In the concordance \LCDM\ models the positions of the acoustic peaks are defined by $\Omega_{\rm m}h^2$ and $\Omega_{\rm bar}h^2$ \citep[see][]{Aubourg2015}. But the propagation of acoustic waves is different in EDE models, as the EDE boosts the Hubble rate around $z_{\rm eq}$ and thus these two cosmological parameters no longer define the BAO peak positions. 

The relative difference between the BAO wiggles in the three cosmologies is better seen in Figure~\ref{fig:BAOwiggles}, where we show the deviations for each linear power spectrum from that without BAO features obtained from the \citep{Eisenstein1998} ``non-wiggle'' P$_{\rm nw}$(k) fitting formula. The BAO shifts among 
the three cosmological models are clearly visible, and systematically shifted towards smaller %frequencies 
wavenumbers by $1.8\%$ for EDE and $4.4\%$ for \LCDM\_EDE with respect to the \LCDM\ BAO positions. This is expected given their corresponding acoustic sound horizon ratios $r_d/r^{\rm fid}_d$, where $r^{\rm fid}_d$ is the sound horizon of our fiducial cosmology, the \LCDM\ model.

\makeatletter{}\begin{figure}
  \centering
\includegraphics[width=0.49\textwidth]
{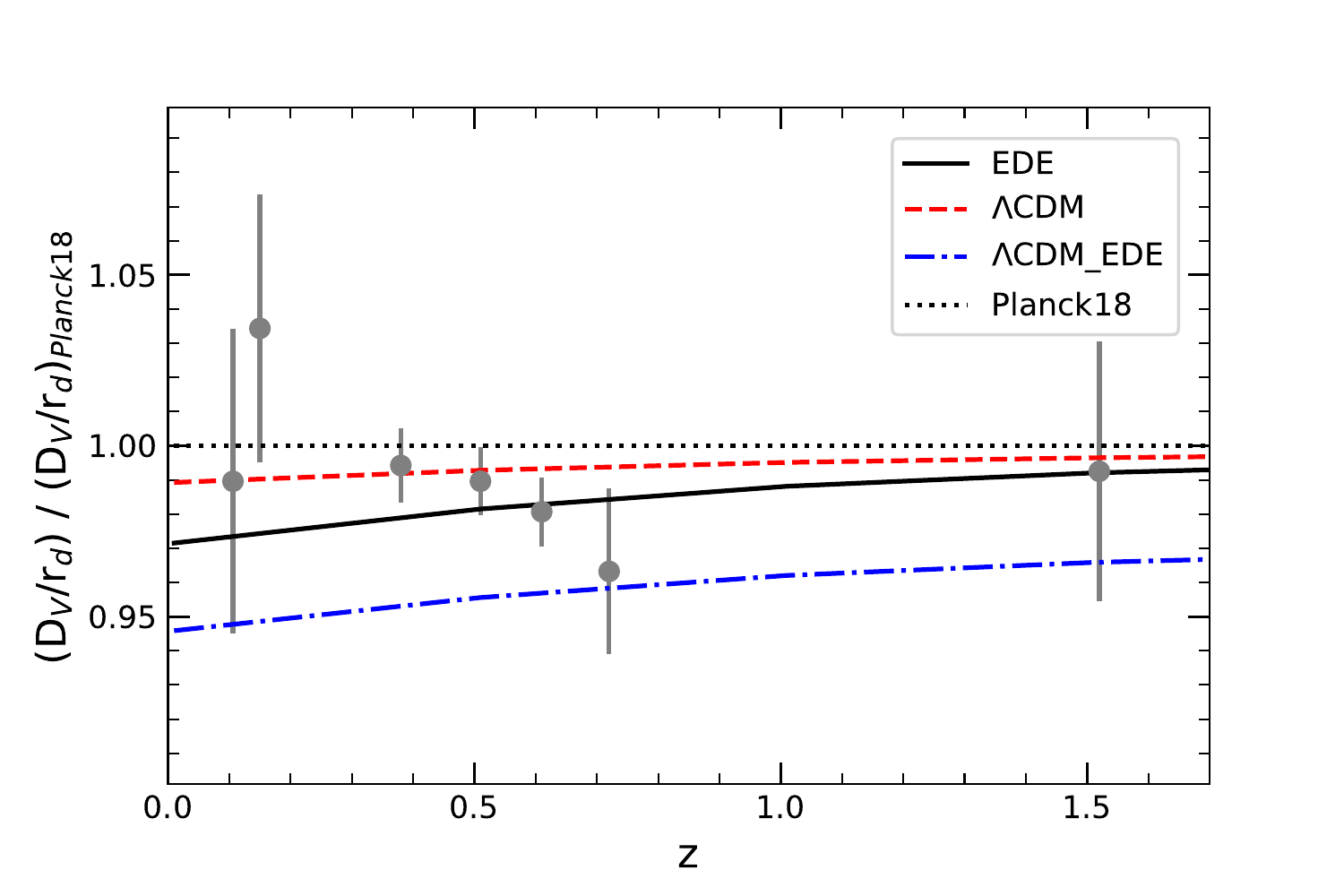}
\caption{Acoustic-scale distance measurements relative to the prediction from Planck TT,TE,EE+lowE+lensing in the base-\LCDM\ model \citep[see Table 1,][]{Planck2018}. The symbols and $1\sigma$ error bars correspond, in increasing redshift order, to the isotropic BAO measurements D$_V$(z)/r$_d$ from the 6dFGRS \citep{Beutler2011}, the SDSS-MGC \citep{Ross2015}, BOSS DR12 LRGs \citep[at $z =$ 0.38, 0.51, and 0.61,][]{Alam2017}, eBOSS DR14 LRGs \citep{Bautista2018} and eBOSS DR14 QSOs \citep{Ata2018}. The curves provide the model predictions from EDE (solid), MultiDark-\LCDM\ (dashed), and \LCDM\ with the same EDE cosmological parameters (dashed-dotted).}
\label{fig:BAOalpha}
\end{figure}

%\footnote{\Vivian{Is there a reason for not showing separately the parallel and perpendicular BAO instead of the spherical average?}}
The BAO position in the spherically averaged two-point clustering statistics, and hence the acoustic-scale distance measurements obtained from large galaxy redshift surveys, are based on the constraints of the stretch or dilation parameter $\alpha$ defined as,
\begin{equation}
\alpha \equiv  \frac{(D_{\rm V}(z)/r_{\rm d})}{(D^{\rm fid}_{\rm V}(z)/r^{\rm fid}_{\rm d})},
\label{eq:alpha}
\end{equation}
where $D_{\rm V}(z)\equiv [cz (1+z)^2 D^2_A H^{-1}(z)]^{1/3}$ is the dilation distance \citep{Eisenstein2005}, $D_A$ is the 
angular diameter distance and $H(z)$ is the Hubble parameter. The stretch
parameter $\alpha$ is measured from the best-fit model to the observed isotropic power spectrum or correlation function on 
the scale range $0.05\,h/{\rm Mpc} \lesssim  k  \lesssim 0.3\,h/{\rm Mpc}$ \citep[see, e.g.,][]{Anderson2014,Ross2015}. The latest and 
more accurate acoustic-scale distance $D_{\rm V}/r_{\rm d}$ measurements, relative to the prediction from 
$Planck$ TT,TE,EE+lowE+lensing (CMB) in the base-\LCDM\ model (i.e., our fiducial MultiDark-Planck13 cosmology, see Table~\ref{table:cosmology}) are 
shown in Figure~\ref{fig:BAOalpha}. 
%We have shown, in increasing redshift order, 
% the currently available isotropic BAO results.
%from 
%the 6dFGRS \citep{Beutler2011}, the SDSS-MGC \citep{Ross2015}, BOSS DR12 LRGs \citep{Alam2017}, eBOSS DR14 LRGs \citep{Bautista2018} 
%and eBOSS DR14 QSOs \citep{Ata2018}. 
The curves in Figure~\ref{fig:BAOalpha} correspond to the model predictions from EDE (solid), MultiDark-\LCDM\ (dashed), 
and \LCDM\ with the same EDE cosmological parameters (dashed-dotted). We conclude that EDE and our \LCDM\ cosmology models both 
agree well with the observations. 

The effect of early dark energy clearly shows up at later epochs, having its maximum 
difference $\sim 2\%$ at $z=0$ as compared to \LCDM. 
The upcoming DESI\footnote{https://www.desi.lbl.gov} and 
Euclid\footnote{https://sci.esa.int/web/euclid} experiments with sub-percent accuracy on the acoustic scale measurements 
will be able to test models such as the EDE one considered in this work.
It is interesting to note that \LCDM\_EDE, despite having the same cosmological 
parameters as EDE but not the same sound horizon scale,  predicts $\alpha$ that differs substantially at all redshifts by 
about $4\%$) (see Figure~\ref{fig:BAOalpha}).

\makeatletter{}\begin{table*}
\begin{center}
\caption{Mean values of the BAO shift and damping at different redshifts obtained from the best-fit $\alpha$ and $\Sigma_{nl}$ parameters of the $\sim 200$ realizations of EDE$_{2B}$ and \LCDM$_{2B}$ real-space power spectra (see Table 2). The damping computed from linear theory $\Sigma^{th}_{nl}$ for each cosmology is also listed for comparison.
}
\begin{tabular}{ c | c | c | c | c | c | c }
\hline\hline  
  &  & \LCDM &  &  & EDE  &  \\
  &  &  &  &  &   &  \\
redshift & $\alpha-1$[\%] & $\Sigma_{nl}$ (Mpc/h) & $\Sigma_{nl}^{th}$ & $\alpha-1$[\%] & $\Sigma_{nl}$ (Mpc/h) & $\Sigma_{nl}^{th}$  
\tabularnewline
  \hline
4.079 &\ \ $ 0.061 \pm  0.020 $ &\ \ $ 2.089 \pm  0.052 $ &\ \  2.101 &\ \ $  0.060 \pm  0.019 $ &\ \ $  2.090 \pm  0.055 $ &\ \  2.170	\\
2.934 &\ \ $ 0.080 \pm  0.022 $ &\ \ $ 2.745 \pm  0.039 $ &\ \  2.703 &\ \ $  0.078 \pm  0.020 $ &\ \ $  2.787 \pm  0.042 $ &\ \  2.791	\\
1.940 &\ \ $ 0.115 \pm  0.024 $ &\ \ $ 3.650 \pm  0.032 $ &\ \  3.584 &\ \ $  0.110 \pm  0.023 $ &\ \ $  3.729 \pm  0.033 $ &\ \  3.699	\\
1.799 &\ \ $ 0.123 \pm  0.024 $ &\ \ $ 3.823 \pm  0.031 $ &\ \  3.756 &\ \ $  0.117 \pm  0.023 $ &\ \ $  3.907 \pm  0.032 $ &\ \  3.876	\\
1.553 &\ \ $ 0.139 \pm  0.025 $ &\ \ $ 4.162 \pm  0.030 $ &\ \  4.095 &\ \ $  0.132 \pm  0.024 $ &\ \ $  4.255 \pm  0.031 $ &\ \  4.224	\\
1.256 &\ \ $ 0.165 \pm  0.027 $ &\ \ $ 4.650 \pm  0.029 $ &\ \  4.589 &\ \ $  0.154 \pm  0.027 $ &\ \ $  4.755 \pm  0.030 $ &\ \  4.731	\\
1.021 &\ \ $ 0.182 \pm  0.026 $ &\ \ $ 5.095 \pm  0.026 $ &\ \  5.063 &\ \ $  0.176 \pm  0.029 $ &\ \ $  5.228 \pm  0.030 $ &\ \  5.215	\\
0.775 &\ \ $ 0.228 \pm  0.032 $ &\ \ $ 5.688 \pm  0.029 $ &\ \  5.656 &\ \ $  0.205 \pm  0.032 $ &\ \ $  5.813 \pm  0.030 $ &\ \  5.820	\\
0.500 &\ \ $ 0.280 \pm  0.036 $ &\ \ $ 6.456 \pm  0.030 $ &\ \  6.465 &\ \ $  0.247 \pm  0.036 $ &\ \ $  6.591 \pm  0.031 $ &\ \  6.641	\\
0.244 &\ \ $ 0.345 \pm  0.042 $ &\ \ $ 7.306 \pm  0.033 $ &\ \  7.377 &\ \ $  0.297 \pm  0.042 $ &\ \ $  7.450 \pm  0.034 $ &\ \  7.560	\\
0.007 &\ \ $ 0.390 \pm  0.043 $ &\ \ $ 8.184 \pm  0.033 $ &\ \  8.354 &\ \ $  0.352 \pm  0.049 $ &\ \ $  8.343 \pm  0.037 $ &\ \  8.536	\\
%\tabularnewline

\tabularnewline
\hline
\end{tabular}
\label{table:baoshifts}
\vspace{-5mm}
\end{center}
\end{table*}

\makeatletter{}\begin{figure}
  \centering
\includegraphics[width=0.49\textwidth]
{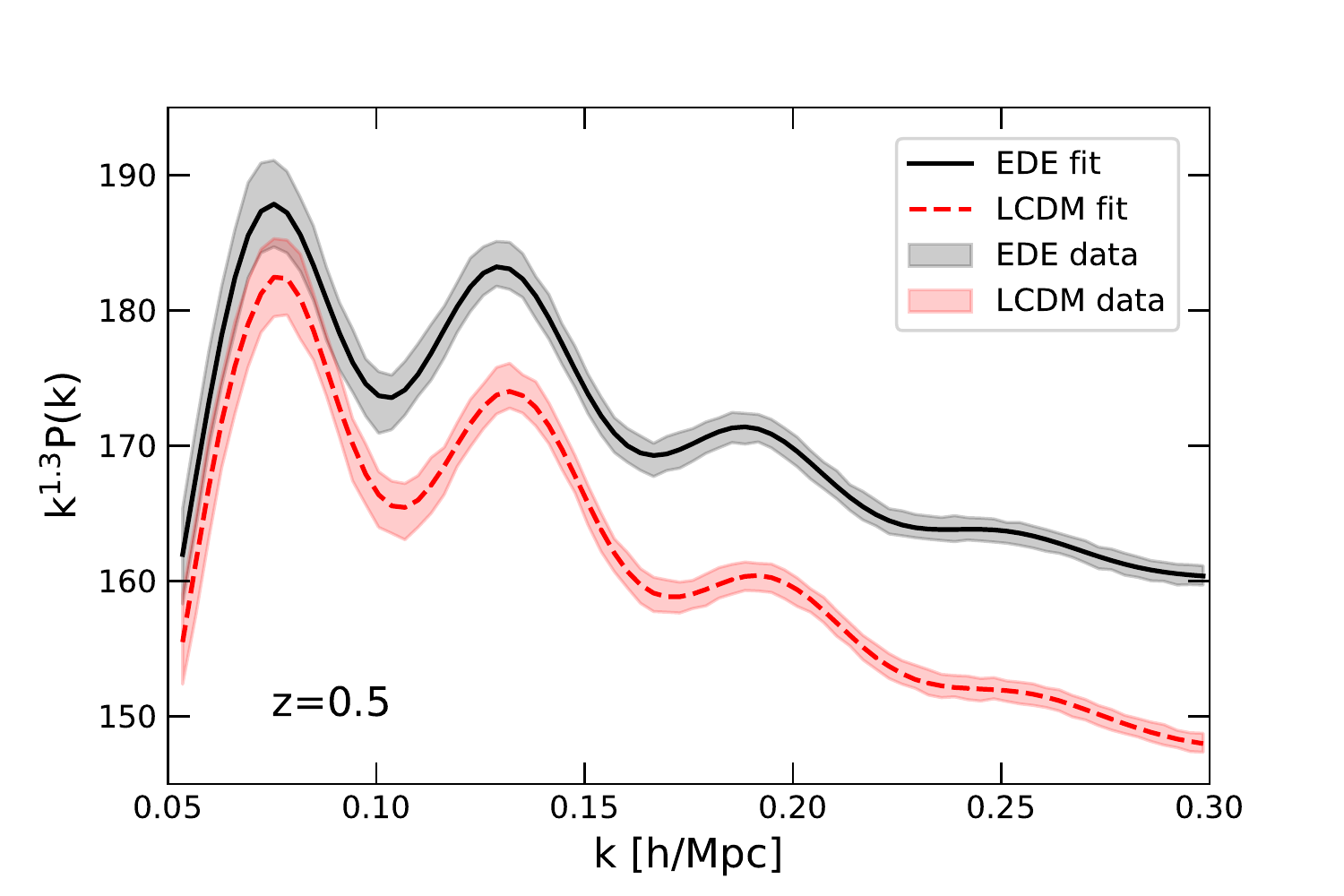}
\caption{Mean, and standard deviation, of the dark matter power spectra at $z=0.5$ obtained from the ensemble of $\sim 200$ EDE and  \LCDM\ GLAM simulations. The solid (dashed) lines correspond to the best-fit model given by Eq.~\ref{eq:BAOmodel} in the wavenumber range $0.05  < k < 0.3$ h Mpc$^{-1}$ for the EDE (\LCDM)\ data.
}
\label{fig:BAOpk}
\end{figure}

\makeatletter{}\begin{figure}
  \centering
\includegraphics[width=0.49\textwidth]
{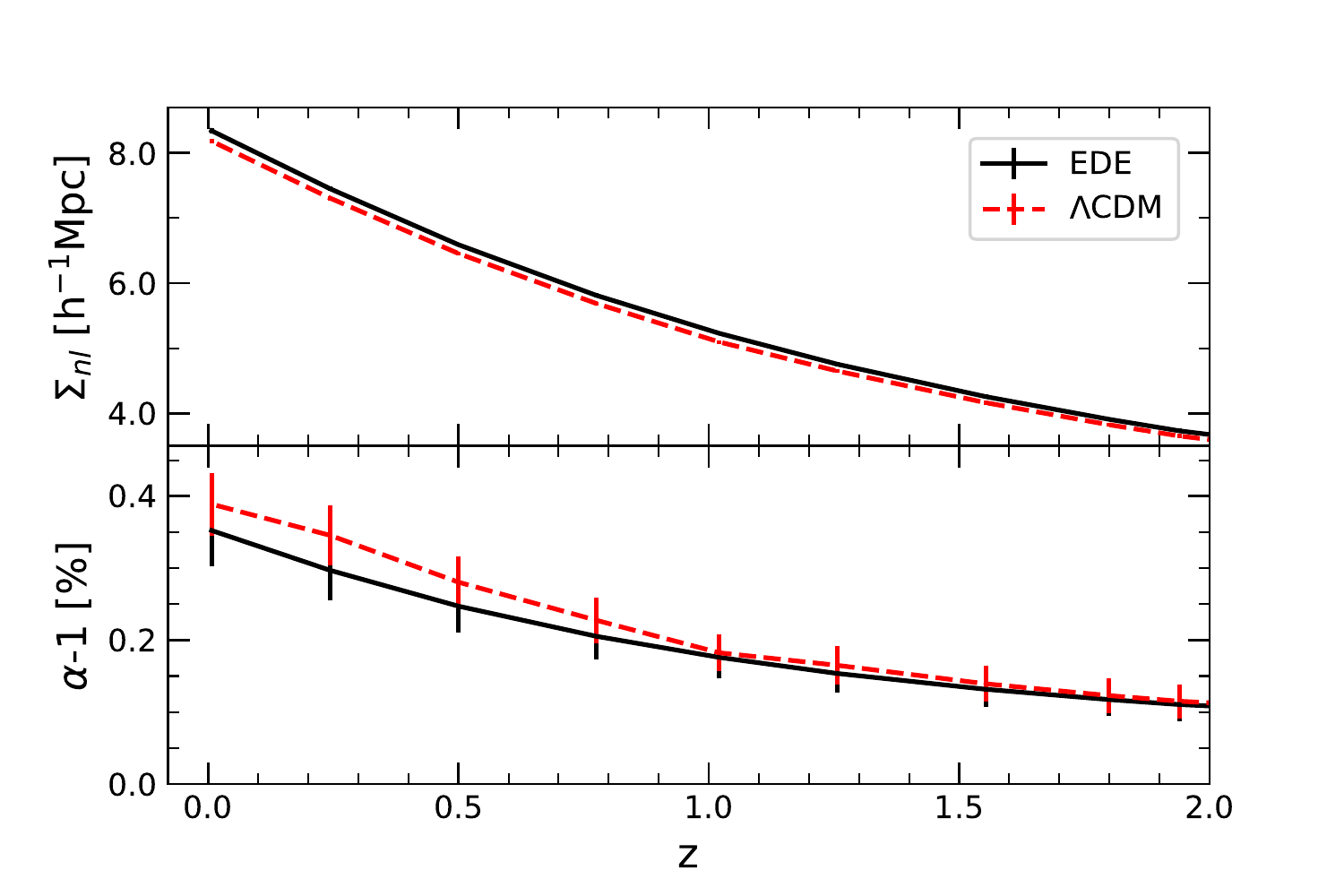}
\caption{Non-linear evolution of the BAO shift (bottom panel) and damping (top panel) for the isotropic dark matter power spectrum in our EDE$_{2B}$ and \LCDM$_{2B}$ simulations. The displayed mean values, and $1\sigma$ uncertainties, of $\alpha$ and $\Sigma_{nl}$, and given in Table~\ref{fig:BAOalpha}, are estimated from the ensemble of individual shifts and damping parameters measured from fitting each of the power spectra, using Eq.~\ref{eq:BAOmodel}, of the EDE and \LCDM\ GLAM simulations.}
\label{fig:BAOdump}
\end{figure}

The acoustic-scale distance measurements up to $z=1.5$ displayed in Figure~\ref{fig:BAOalpha} include density-field reconstruction of the BAO feature, which
is used to partially reverse the effects of non-linear growth of structure formation \citep[see][]{Anderson2012,Padmanabhan2012}. The shape of
the linear matter power spectrum $P(k)$ is distorted by the nonlinear evolution of density fluctuations, redshift distortions and
galaxy bias even at large-scales $k < 0.2 \, \kMpch$. As mentioned above, the shift parameter $\alpha$ yields the relative position of the 
acoustic scale in the power spectrum (or two-point correlation function) obtained from the data (or simulations) with respect to the adopted $P(k)$ model. 

Here we study the non-linear shift and damping of acoustic oscillations up to redshift $z=4$ for dark matter in our ensemble of $\sim 200$ EDE and \LCDM\ GLAM $N-$body simulations. Figure~\ref{fig:BAOpk} shows the spherically-averaged power spectra at $z=0.5$ in real-space drawn for both cosmologies in the domain of the BAO features. We measure the shift of the BAO relative to linear theory by following a similar methodology as that presented in \citet[][]{Seo2008}, and implemented in \citet{Anderson2014} to measure the BAO stretch parameter in the BOSS data. The non-linear dark matter power spectrum with wiggles is modeled by damping the acoustic oscillation features of the linear power spectrum assuming a Gaussian with a scale parameter $\Sigma_{\rm nl}$ which accounts for the BAO broadening due to nonlinear effects \citep[e.g.][]{Eisenstein2007}. We use the functional form:

\begin{equation} \label{eq:BAOmodel}
\begin{split}
P(k) = P_{\rm sm}(k) \left[1 + \left(\frac{P_{\rm lin}(k/\alpha)}{P_{\rm nw}(k/\alpha)} - 1\right) e^{-\frac{1}{2}(k/\alpha)^2 \Sigma_{nl}^2} \right] \, , 
\end{split}
\end{equation}
where $P_{\rm lin}$ is the linear power spectrum generated with CAMB for each cosmology model, and $P_{sm}$ is the smooth "BAO-free" power spectrum modelled as $P_{sm} = P_{\rm nw}(k) + A(k)$ with $P_{\rm nw}(k)$ being the "de-wiggled" \citep{Eisenstein1998} spectrum template and $A(k)$ accounting for the non-linear growth of the broad-band matter power spectrum expressed in the form of simple power-law polynomial terms $A(k) = a_1 k + a_2 + A_3/k + A_4/k
^2 + A_5/k^3$ \citep{Anderson2014}. The shift and damping of the acoustic oscillations, measured by $\alpha$ and $\Sigma_{nl}$, are considered free parameters in our model. 
 
We then perform the fit of the power spectrum $P(k)$ drawn from each of our GLAM simulations over the wavenumber range $0.05  < k < 0.3$ h Mpc$^{-1}$ for several redshifts. The solid (dashed) line in Figure~\ref{fig:BAOpk} corresponds to the best-fit power spectrum model given by using Eq.~\ref{eq:BAOmodel} for the EDE (\LCDM) simulation data. The shift and damping of the BAO features in both cosmologies is similar as can be seen from the plot. The mean values, and $1\sigma$ uncertainties, of the $\alpha$ and $\Sigma_{nl}$ parameters obtained from the best-fit to each of the  EDE and \LCDM\ GLAM power spectra are provided in Table~\ref{table:baoshifts} up to $z=4$. The nonlinear damping
estimated from perturbation theory\footnote{The broadening and attenuation of the BAO feature is exponential, as adopted in our model given in Eq.~\ref{eq:BAOmodel}, with a scale $\Sigma^{th}_{nl}$ computed following \cite{Crocce2006, Matsubara2008}, i.e. $\Sigma^{\rm th}_{\rm nl}=\left[\frac{1}{3\pi^2}\int P_{\rm lin}(k)dk\right]^{1/2}$.} for each cosmology is also listed, and shows a remarkable agreement better than $2\%$ over all redshifts with that measured from our model fits to the simulation data.

Our shift results for the acoustic scale towards larger $k$, relative to the linear power spectrum, and damping values obtained from our analysis are in good agreement with previous works for \LCDM\ \citep[e.g.][]{Crocce2008,Seo2010,Prada2016}. Figure~\ref{fig:BAOdump} demonstrates that the non-linear evolution of the BAO shift (bottom panel) and damping (top panel) for the isotropic dark matter power spectrum in both EDE and \LCDM\ cosmologies display small differences, with the BAO features being less affected by the non-linear growth of structure formation.  Moreover, \citet{Bernal2020} shows that the \LCDM-assumed templates used for anisotropic-BAO analyses can be used in EDE models as well.

\begin{figure}
  \centering
\includegraphics[width=0.49\textwidth]
{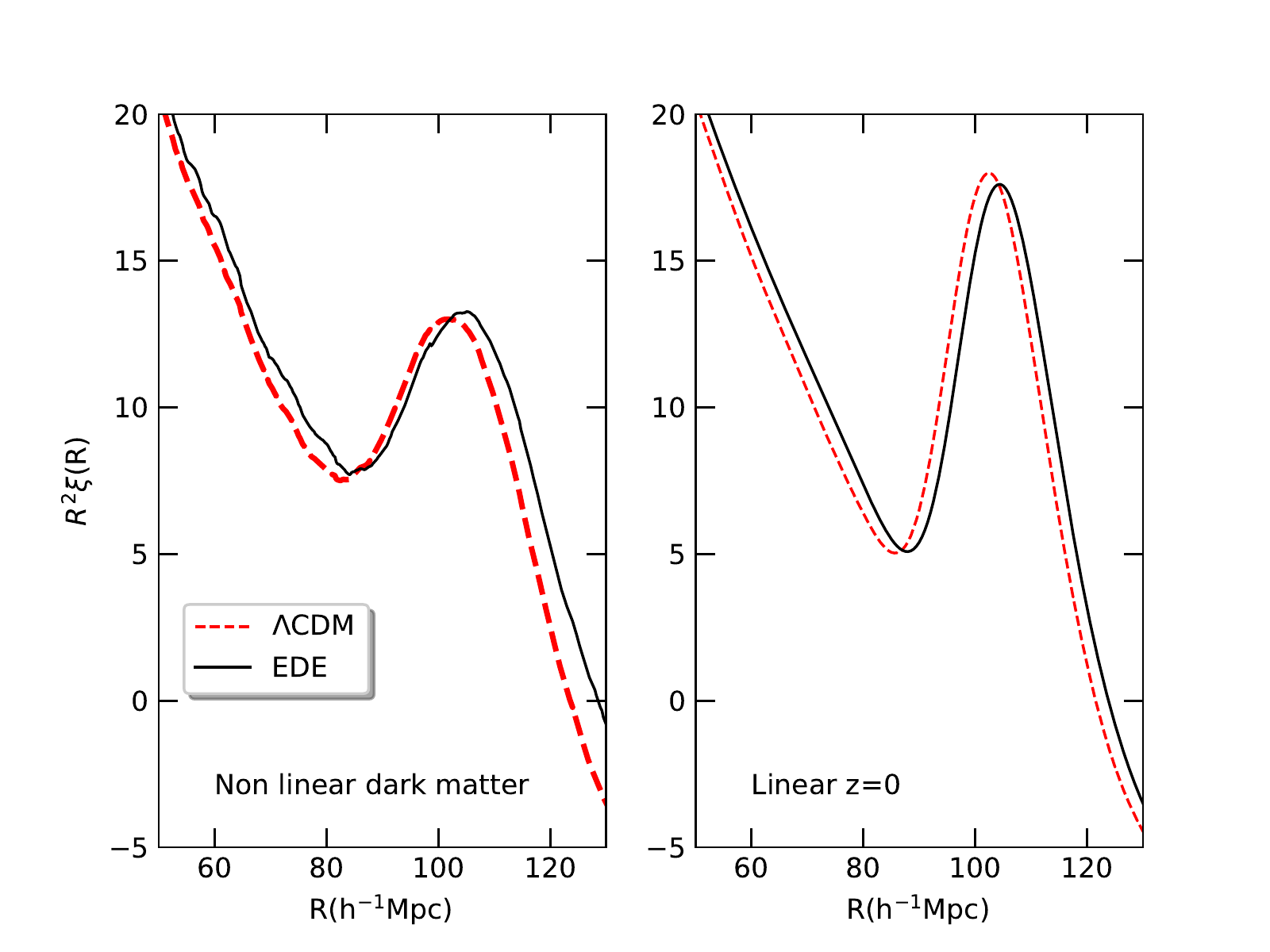}
\caption{{\it Right panel:} Linear correlation function of dark matter
  at $z=0$ on large scales. We plot the correlation function $\xi(R)$
  scaled with $R^2$ to remove the main trend of the correlation
  function. The correlation function in the EDE model is slightly shifted
  by $\sim 2\%$ to larger radii as compered with the \LCDM\ model. {\it
    Left panel:} Nonlinear correlation function
    at $z=0$. As compared to the
  linear $\xi(R)$, the BAO peak in the nonlinear regime slightly shifts
  to smaller values and becomes wider with smaller amplitude --
  effects that are well known and well understood. Nonlinear effects do
  not reduce differences between EDE and \LCDM\ models. }
\label{fig:CorrelationDM}
\end{figure}

A summary of our BAO results can also be shown in configuration space. In Figure~\ref{fig:CorrelationDM} we see that the BAO peak in the EDE linear correlation function (right panel) is slightly shifted by $\sim 2\%$ to larger radii as compared with the \LCDM\ model, as expected from their different values of the sound horizon scale at the drag epoch. The impact of non-linear evolution broadens the BAO peaks but it does not reduce the shift differences between EDE and \LCDM.

\makeatletter{}

\section{Halo abundances}
\label{sec:abundance}

\makeatletter{}\begin{figure}
  \centering
\includegraphics[width=0.52\textwidth]
{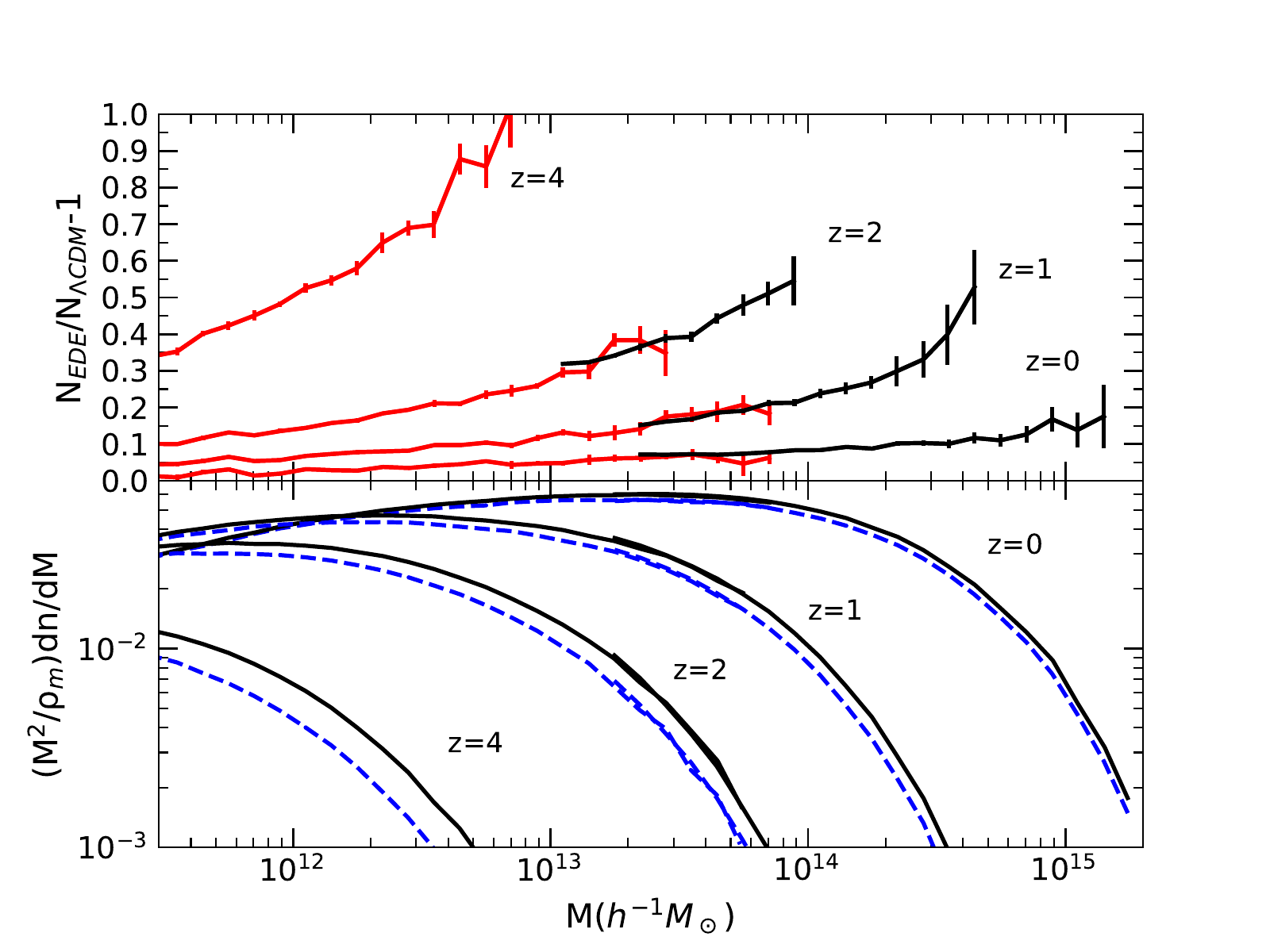}
\caption{Halo mass function at redshifts $z=0-4$. Full curves in the
  bottom panel are for the EDE simulations and dashed curves are for the
  \LCDM\ simulations. The smaller box and better resolution simulations EDE$_{0.5}$ and
  \LCDM$_{0.5}$ are used for masses below $M\lsim 10^{14}\Mpch$. They
  are shown as red curves in the top panel. Larger box and lower
  resolution simulations EDE$_{2A}$ and \LCDM$_{2A}$ (black curves in
  the top panel) are used for massive halos with
  $M\gsim 2\times 10^{13}\Mpch$. At $z=0$ halo abundances are very
  similar for the models: EDE predicts $\sim 10\%$ more of the most
  massive clusters $M\approx 10^{15}\Msunh$ and 1\%-2\% more of
  galaxy-size halos with $M\approx 10^{12-13}\Msunh$. The differences
  in abundances increase substantially with the redshift.}
\label{fig:HMS}
\end{figure}

\makeatletter{}\begin{figure}
  \centering
\includegraphics[width=0.52\textwidth]
{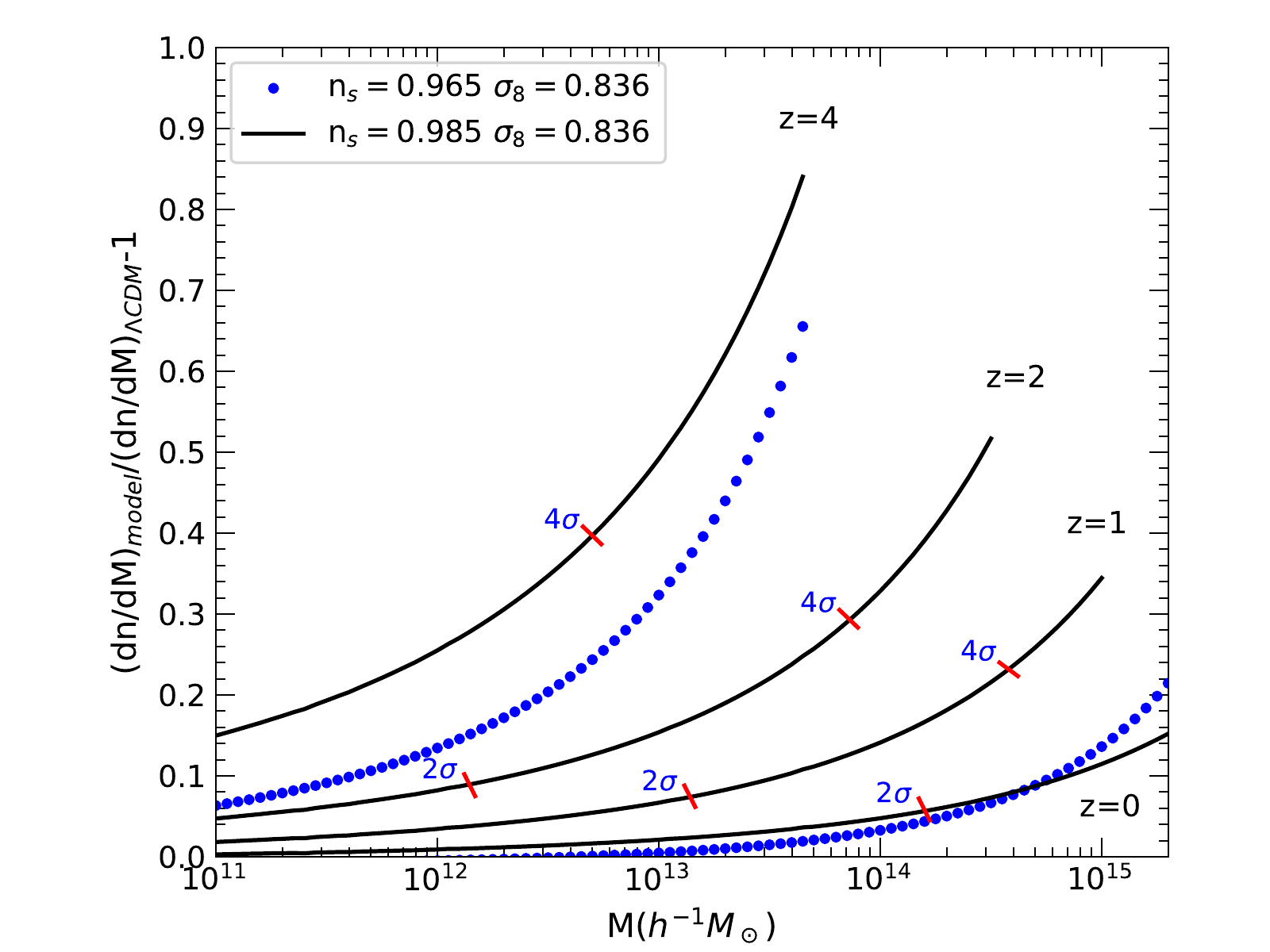}
\caption{Analytical estimates of the ratio of halo abundances of
  different models relative to the abundance in the \LCDM\
  model. The \citet{Despali2016} approximation is used to make the
  predictions. Dotted curves for $z=0$ and $z=4$
  show effects due to the increase of just
  normalization from $\sigma_8=0.820$ in \LCDM\ to $\sigma_8=0.836$
  (as in the EDE model). The full curves are for the model with
  increased slope $n_s=0.985$ (as compared to $n_s=0.965$ in \LCDM)
  and the increased $\sigma_8=0.836$. Small lines mark positions of
  peaks of given $\nu$ height. As expected, the curves start to
  go up steeply when halos become high peaks of the Gaussian
  field. The analytical models qualitatively explain the main
  differences between the EDE and \LCDM\ models, although they
  underpredict the magnitude of the real differences observed in
  Figure~\ref{fig:HMS}. }
\label{fig:HMStoy}
\end{figure}

To study halo mass functions we use simulations with 500\Mpch\ and
1000\Mpch\ boxes and mesh size $N_g=7000$. Simulations with larger
2\Gpch\ boxes have lower mass and force resolutions -- not sufficient
for analysis of galaxy-mass halo abundances.

Halos in simulations were identified with the Spherical Overdensity
halofinder BDM \citep{Bolshoi,Knebe2011} that uses the virial
overdensity definition of \citet{BryanNorman}. The resolution was not 
sufficient for identifying subhalos, so only distinct halos are
studied.

Figure~\ref{fig:HMS} shows the halo mass function at different
redshifts. The EDE model predicts more halos at any redshift, but the
difference is very small at $z=0$: a 10\% effect for very massive
clusters $M\approx 10^{15}\Msunh$ and just 1\% for Milky Way-mass
halos with $M=10^{12}\Msunh$. These differences hardly make any impact
on predicted statistics of galaxies and clusters with observational
uncertainties and theoretical inaccuracies being larger than differences
in halo abundances.

The situation is different at larger redshifts: the number of halos in EDE
is {\it substantially} larger than in \LCDM. For example, the EDE model
predicts about 50\% more massive clusters of mass
$M=(3-5)\times 10^{14}\Msunh$ at $z=1$.  The differences increase even more at
larger redshifts. For example, the EDE model predicts almost
twice more galaxy-size halos with $M>3\times 10^{12}\Msunh$ at
redshift $z=4$.  These are interesting predictions that can
potentially be tested by comparing with abundances of high redshift
$z\gsim 1$ clusters of galaxies
\citep[e.g.,][]{Bayliss2014,Gonzales2015,SPT2019}, abundances of
massive galaxies and black holes at $z>4$
\citep[e.g.,][]{HaimanLoeb,Stefanon2015,Behroozi2018,Carnall2020}, 
and clustering of high-redshift galaxies \citep{Harikane16,Harikane18,Endsley2020}.

Another consequence of the increased mass function in EDE is earlier collapse
times.  More halos in EDE at higher redshifts implies that halos of a
given mass $M$ form earlier in the EDE model.  Because the Universe is
denser at those times, so are the halos. At later times the accretion of
dark matter onto the halo gradually builds the outer halo regions
resulting in increasing halo concentration
\citep[e.g.,][]{Bullock2001}.  Thus denser central regions in EDE models
should lead to more concentrated halos.

At first sight our results on the halo mass functions are puzzling. Halo
mass functions are defined by the amplitude of perturbations
$\sigma(M,z)$. However, the normalization of the perturbations
$\sigma_8$ is just 2\% different in the EDE model. Why do we see large
deviations in the halo abundances? The evolution of the mass function
is defined by the growth rate of fluctuations, which in turn is
defined by $\Omega_m$, which is nearly the same for EDE and \LCDM\
models. In this case why do we see large evolution of the differences between the
models? In order to have some insights on the issue, we use analytical
estimates of the halo mass function that allow us to change parameters and see their effects.

Specifically, we use the \citet{Despali2016} approximation for virial halo
mass function at different redshifts. By itself the approximation is
not accurate enough to reliably measure the differences between EDE
and \LCDM\ models.  However, it is good enough to study trends and to probe
effects of different parameters.

According to the theory \citep[e.g.,][]{Bond1991,ST}, the halo mass
function $n(M,z)$ is a function of $\sigma(M,z)$ -- the $rms$ of the
linear density field smoothed with the top-hat filter of radius $R_f$
corresponding to the average mass $M$ inside a sphere of radius $R_f$:
$M = (4\pi/3)\rho_mR_f^3$.  Spherical fluctuations that in the linear
approximation exceed a density threshold $\delta_{\rm cr}\approx 1.68$
in the real nonlinear regime collapse and form dark matter halos.  The
halo mass function
\begin{equation}
  \frac{dn}{dM} = f(\sigma) \left[\frac{\Omega_m\rho_{\rm crit}}{M^2}\right]\frac{d\ln\sigma}{d\ln M}
\end{equation}
can be
written in a form that depends mostly on one parameter -- the relative
height of the density peak $\nu$ defined as:
\begin{equation}
\nu = \frac{\delta_{\rm cr}}{\sigma(M,z)}.
\end{equation}
There are different approximations for function $f(\sigma)$. We start
with the Press-Schechter approximation because it is easy to see the main
factors defining the mass function:
\begin{equation}
 \left(\frac{M^2}{\Omega_m\rho_{\rm crit}}\right)\frac{dn}{dM} = 
\sqrt\frac{2}{\pi}\nu \exp\left(-\frac{\nu^2}{2}\right)\frac{d\ln\sigma}{d\ln M}.
\label{eq:PS}
\end{equation}
When $\nu$ is small ($\nu\lsim 1$), the Gaussian term is close to
unity, and the amplitude of the mass functions is linearly
proportional to $\nu$, which, in turn, is
inversely proportional to the
normalization $\sigma_8$. This explains why the EDE mass function is just
$\sim 1\%-2\%$ larger than in \LCDM\ at small $M$ and at $z=0$: we are
dealing with small $\nu$ peaks of the Gaussian density field. As mass
increases, the {\it rms} of fluctuations $\sigma(M)$ decreases, and
eventually $\nu$ becomes large.  In this case the Gaussian term
dominates, and we expect a steep decline of $dn/dM$. In this regime
the ratio of mass functions is equal to $\approx\exp(\alpha\nu^2)$,
where
$\alpha =(\sigma_{8,\rm EDE}/\sigma_{8,\rm \LCDM})-1\approx 0.02$. For
example, for $4\sigma$ fluctuations $\nu = 4$, we expect a
$\sim 40\%$ difference.  In other words, for high-$\nu$ peaks a small
change in the amplitude of fluctuations produces a very large change
in the halo abundance. This is exactly what we see in
Figure~\ref{fig:HMS} at large redshifts.
 
In practise, we use a better approximation for the halo mass function
provided by \citet{Despali2016}. We find that the approximation is
very accurate at low redshifts with the errors less than $\lsim 3\%$
for masses $M_h>10^{12}\Msunh$. However, the errors increase with the
redshift, becoming $\approx 12\%$ at $z=4$ for the \LCDM\ model. The errors
also depend on cosmology: at $z=4$ the error for the EDE model is
$\approx 30\%$. While not very accurate, the approximation can be used
for qualitative analysis.

We are mostly interested in effects of modification of the amplitude
$\sigma_8$ and in changes of the slope of the spectrum $P(k)$. For the
base model we use \LCDM\ with $\sigma_8=0.820$. We start with increasing the amplitude to the
same value $\sigma_8=0.836$ as in the EDE model. When doing this, we take the same
shape of spectrum as in \LCDM\ and increase the normalization. Dotted curves in Figure~\ref{fig:HMStoy} show
how the mass function changes due to the increased $\sigma_8$. As
expected, the high-$\sigma_8$ model has more halos and the difference
increases with mass and with the redshift. However, the shape of the
mass function ratios is too steep as compared with the $N$-body
simulations. Compare, for example, the $z=0$ curves in
Figure~\ref{fig:HMS} and Figure~\ref{fig:HMStoy}. Also, the magnitude of the
effect is much smaller at $z=4$ as compared with what it should be.

Now we also change the slope of the power spectrum from $n_s=0.965$ to
the same value $n_s=0.985$ as in the EDE model while keeping the same high 
normalization $\sigma_8$. Because the radius $R_f=8\Mpch$ of the
top-hat filter in the $\sigma_8$ definition was chosen such that the
abundance of massive clusters with $M\approx 10^{15}\Msunh$ should
stay approximately constant, keeping the same $\sigma_8$ means that
cluster abundance does not change much. At the same time a steeper
slope of $P(k)$ means that the amplitude of fluctuations increases for
small halos. As the result, the full curves for tilted and
high-$\sigma_8$ models in Figure~\ref{fig:HMStoy} are flatter producing
more halos with small mass.

In Figure~\ref{fig:HMStoy} we also mark positions of peaks of given $\nu$
height. As expected, the curves start to steeply go up when halos
become high peaks of the Gaussian field.

In summary, the EDE model predicts quite similar ($1-10\%$) halo
abundance as \LCDM\ at low redshifts, significantly increasing
%and much larger number (up to a factor of 2) 
at higher redshifts. Most of the increase is due to the change
$\Delta n_s=0.02$ in the slope of the power spectrum with the
increase in $\sigma_8$ playing an additional role. These results are well
understood in the framework of the theory of the halo mass function,
although the analytical approximation by \citet{Despali2016} fails to
reproduce the results accurately with errors up to $\sim 30\%$ being
redshift- and model-dependent.

\section{Halo abundances and clustering at high redshifts}
\label{sec:abundanceZ}
 Results discussed in the previous section show a remarkable increase with 
 redshift in halo abundances in the EDE model (relative to the \LCDM\ model).
 Here, we study 
 predictions for even larger redshifts. We focus on two issues: (a) the abundance of 
 small halos at the epoch of recombination ($z=6-10$) and (b) the clustering of
 halos at $z=4-6$ that are plausibly measurable with \textit{JWST} \citep{Endsley2020}.
 
\makeatletter{}\begin{figure}
  \centering
\includegraphics[width=0.52\textwidth]{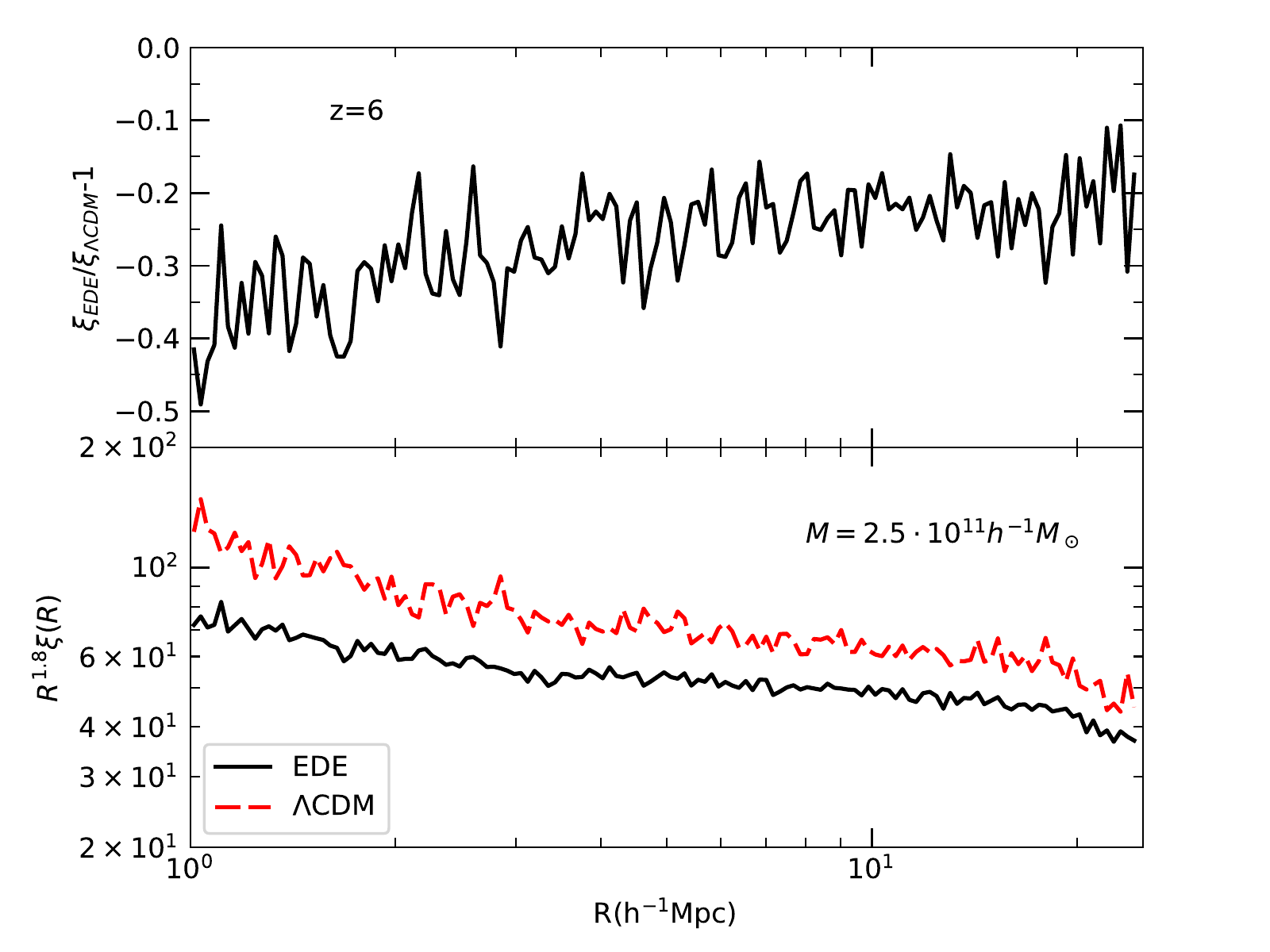} 
\includegraphics[width=0.52\textwidth]{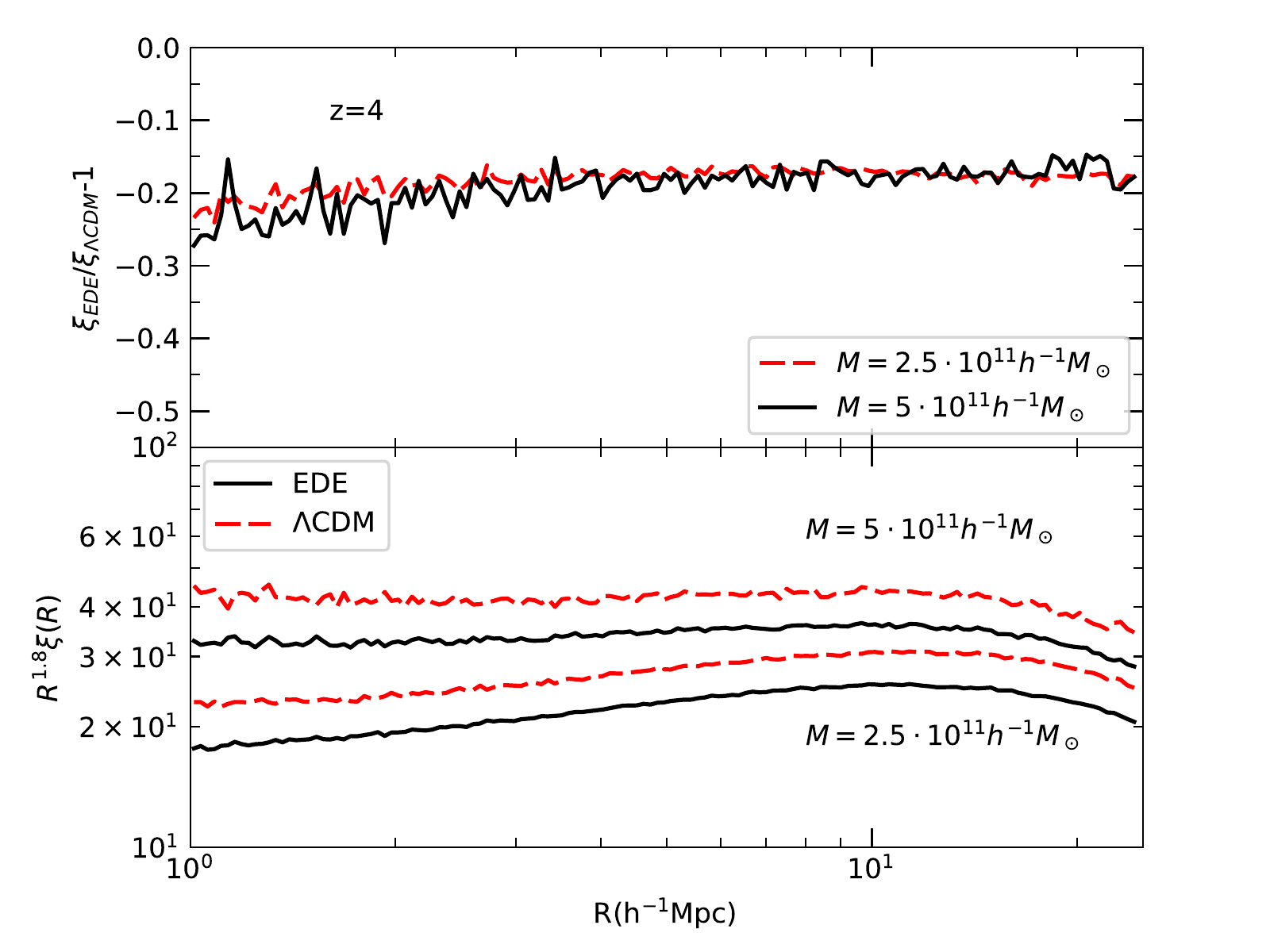}
\caption{Correlation functions of dark matter halos at redshift $z=6$ (top panels) and $z=4$ (bottom panels). 
Halos above virial masses indicated in the plots 
were used to find the correlations and their ratios. In the distance range $R=(1-20)\Mpch$ the correlation 
functions are well approximated by a power-law  $\xi(R)\propto r^{-1.8}$.
At each redshift halos in the EDE model are less clustered by $\sim (10-30)\%$ than halos with the same 
mass cut in the \LCDM model -- an unexpected result considering that the dark matter in the EDE model is more clustered.}
\label{fig:XiZ6}
\end{figure}
 
 We make additional  simulations 
 using smaller simulation boxes of $50\Mpch$ and $250\Mpch$ with $2000^3$ particles and 
 force resolutions of $7\kpch$ and  $36\kpch$ correpondingly, which is substantially  better than in the $EDE_{0.5}$ and $\LCDM_{0.5}$ 
 simulations. In addition to our EDE and \LCDM\ models, we also run a  
 simulation $\LCDM_{\rm low}$ with the same parameters as the \LCDM model\, but 
 with lower amplitude of fluctuations $\sigma_8=0.75$ that is motivated 
 by weak-lensing results \citep[e.g.,][]{Hamama2020}.
 %(\textbf{citations}). 
 Because of the smaller box size, the improved 
 mass resolution of these simulations ($m_p=1.3\times 10^6\Msunh$ and $m_p=1.6\times 10^8\Msunh$) allows 
 us to study halo abundances and halo clustering for halos with masses as
 low as $\sim 10^9 - 10^{10}\Msunh$.

%Historically it was always difficult to explain 
The abundance and clustering of such low-mass halos is particularly relevant for understanding reionization.  The Universe was
re-ionized between $z=6-10$ \citep[e.g.,][]{Madau2014}, and it is generally accepted that 
the observed population of relatively bright star-forming galaxies ($M_\mathrm{UV}<-17$; $M_h > 10^{10}\Msunh$) cannot 
provide enough ionizing photons \citep{Paardekooper2015,Robertson2015,Finkelstein2019}.
However,  fluxes from fainter galaxies may be sufficient \citep{Robertson2015,Yung2020}.
The predicted ionizing flux of UV radiation depends on three factors: efficiency 
of star formation (especially in low-mass halos), abundance of halos of 
different masses at the epoch of re-ionization, and the escape fraction of photons. 
Theoretical estimates \citep{Finkelstein2019,Yung2020} indicate that about 50-60\% of ionizing
photons were produced by (but not necessarily escaped from) galaxies hosted in
halos with masses $M_h=10^{10}-10^{12}\Msunh$. These estimates are based on halo 
abundances in the standard \LCDM\ model. Most of the radiation came from galaxies hosted
by halos with mass $M_h>10^9\Msunh$ \citep{Barkana2001,Finkelstein2019}.

We find that the trend of increasing halo abundance ratios persists during the epoch of re-ionization at 
redshifts $z=6-10$. For example, at $z=7.5$ the EDE model predicts 
$1.8$ times more halos with masses larger than $M=10^{10}\Msunh$ as 
compared with \LCDM. The difference with $\LCDM_{\rm low}$ is even 
more striking: there are more than 3.7 times more halos above that mass cut in the EDE 
as compared with the $\LCDM_{\rm low}$ model. At $z=10$, the $\LCDM_{\rm low}$ 
model has 8.3 times fewer halos with $M>5\times 10^{9}\Msunh$ as compared with the EDE model.

Thus, with other parameters fixed, the EDE model would predict a factor of $\sim 1.5-2$ larger ionizing fluxes as compared
with the \LCDM\ model. Reducing the fluctuation amplitude to $\sigma_8 =0.75$ would result in reduction
of fluxes by a factor of 3-5 compared with \LCDM, 
which would be problematic.

Clustering of high-redshift galaxies  is potentially an 
interesting 
%statistics 
way to distinguish different cosmological models. Ground observations with the Hyper
Suprime-Cam (HSC) Subaru telescope  \citep[e.g.,][]{Harikane16,Harikane18} and future measurements of large
samples of galaxies at $z=4-6$ with JWST \citep{Endsley2020} will bring an opportunity to combine galaxy
clustering with abundances as a probe for halo masses and merging rates. Theoretical estimates 
indicate that the observed galaxies at those redshifts should be hosted by halos with masses in the range $M=10^{10}-10^{12}\Msunh$ \citep{Harikane16,Harikane18,Endsley2020}.

As we saw earlier, the EDE model predicts stronger dark matter clustering and larger halo abundances at
high redshifts as compared with the \LCDM\ model. Thus, one would naively expect that halos--and 
galaxies hosted by those halos--should also be more clustered. However, our simulations show that this is not the case. Here, we 
use distinct halos in the EDE$_{0.5}$ and $\Lambda$CDM$_{0.5}$ simulations to study clustering of halos
with masses $M =(2-5)\times 10^{11}\Msunh$. Figure~\ref{fig:XiZ6} shows the results for halos at $z=6$ (top panels) and $z=4$ 
(bottom panels). Note that the smallest radius plotted, $R=1\Mpch$, is significantly larger than the virial radii of these halos. Thus, 
the radii presented in the Figure are well in the domain of the two-halo term and are well resolved by the simulations.

The halo correlation functions at those redshifts and radii are nearly power-laws, $\xi(R)\approx (R/R_0)^\gamma$, with 
slopes $\gamma \approx -1.7-1.9$,
which is similar to the slope of Milky-Way-mass halos at $z=0$. The amplitudes of clustering ($R_0$) at high redshifts are
remarkably large. For example, at $z=6$ and $M>2.5\times 10^{11}\Msunh$, the clustering scale is $R_0=9.9\Mpch$ for \LCDM\ and 
$R_0=8.8\Mpch$ for the EDE model.

Somewhat unexpectedly, the clustering of halos in the EDE model is smaller than in  \LCDM\ in spite of the fact that the dark matter
is more strongly clustered in EDE. The differences depend on redshift and halo mass, but those dependencies are weak. Overall, at the
same mass cut, halos in EDE have correlation functions $\approx 20-30\%$ smaller. When we select halos with the 
same cumulative number density, the differences become even smaller ($\approx 10\%$).  This suggests that measuring clustering at fixed galaxy number density will not be a strong test of EDE.  Instead, other mass-sensitive measures (e.g., satellite kinematics or redshift-space distortions) may be more successful probes at high redshifts.

\makeatletter{}\section{Summary and Discussion}
\label{sec:concl}

There are two main tensions between the standard \LCDM\ cosmology and local observations, the Hubble tension and the $S_8$ tension.  The EDE model considered here resolves the Hubble tension, which is that Planck-normalized \LCDM\ predicts a value of the cosmological expansion rate that is smaller than local measurements by as much as 6$\sigma$. Such a large discrepancy is unlikely to be a statistical fluke. And it is probably not due to systematic errors because it is seen in different kinds of measurements, in particular Cepheid-calibrated SNe Ia giving $h = 0.674\pm0.006$ \citep[][the SH0ES team]{SH0ES} and strong-lens time delays giving $h=0.733\pm0.018$ \citep[][the H0LiCOW team]{Wong2020}. 

As we discussed in the Introduction, this Hubble tension can be resolved by adding a maximum of 10\% of dark energy to the energy density of the universe for a brief period around the end of the radiation domination era at redshift $z \approx 3500$ \citep[][Figure 1] {SmithEDE2020}.  
As we also discussed in the Introduction, this EDE model does not exacerbate the relatively small ($\sim 2\sigma$) $S_8$ tension in standard ΛCDM.
    
    In this paper on the EDE model we have focused on the nonlinear effects on halo abundance and clustering, including the baryon acoustic oscillations.  
On large scales, the small differences between the linear power spectra of standard \LCDM\ and our EDE model are mostly due to the different $\sigma_8$ values.  But on smaller scales the linear theory differences become larger because of the slightly larger slope $n_s$ of the EDE primordial power spectrum. Similar effects are expected in other EDE models that are motivated by resolving the Hubble tension \citep[e.g.,][]{Argawal2019,Lin2019}.

In this paper we have explored the nonlinear implications of the \citet{SmithEDE2020} EDE model using a large suite of cosmological $N$-body simulations.   When nonlinear effects are taken into account, standard \LCDM\ and EDE differ by only about 1-10\% in the strongly non-linear regime $k\gsim 1\kMpch$ at low redshift.  On the larger scales of the baryon acoustic oscillations (BAOs), in linear theory the peaks are shifted to smaller wavenumbers by about $2$\% as a consequence of the different value of the sound horizon scale at the drag epoch. Nonlinear effects broaden and damp the BAO peaks, but the $\sim 2$\% shift to larger physical scales is robust.  As Figure~\ref{fig:BAOalpha} shows, both standard \LCDM\ and the EDE model in good agreement with all the presently available acoustic-scale distance measurements. DESI and Euclid measurements will soon be able to test such predictions more stringently.  

The mass function of distinct dark matter halos (those that are not subhalos) is very similar to that of standard \LCDM\ at $z=0$, but the number of halos in EDE becomes substantially larger at higher redshifts.  An analytic analysis shows that the number of halos increases a lot compared to \LCDM\ when they correspond to fluctuations with high amplitude $\nu$, where the Gaussian term in the mass function dominates.  The increase in the number of rare cluster-mass halos at $z \gsim 1$ is mainly due to the increase in $\sigma_8$ in the EDE model, while the increase in $n_s$ causes a further increase in the number of galaxy-mass halos at high redshift.

Our $N$-body simulations of the nonlinear evolution of the EDE model show that its power spectrum and halo mass functions agree within a few percent with those of standard \LCDM\ at redshift $z=0$, so the successful predictions of standard \LCDM\ at low redshifts apply equally to the EDE cosmology.  However, the EDE model predicts earlier formation of dark matter halos and larger numbers of massive halos at higher redshifts.  This means that halos of the same mass will tend to have higher concentrations.  However, they will not have increased clustering.  These predictions will be tested by upcoming observations, with all-sky cluster abundances being measured by the eROSITA X-ray satellite, and the abundance and clustering of high-redshift galaxies to be measured especially by JWST \citep[e.g.,][]{Endsley2020}.

Higher resolution simulations will be needed for more detailed comparisons with observations. We leave those to future work.

\textbf{}

\section*{Acknowledgements}
AK and FP thank the support of the Spanish Ministry of Science funding grant PGC2018-101931-B-I00. This work used the skun6@IAA facility managed by the Instituto de Astrof\'{\i}sica de Andaluc\'{\i}a (CSIC). The equipment was funded by the Spanish Ministry of Science EU-FEDER infrastructure grant EQC2018-004366-P. MK acknowledges the support of NSF Grant No.\ 1519353, NASA NNX17AK38G, and the Simons Foundation. TLS acknowledges support from NASA (though grant number 80NSSC18K0728) and the Research Corporation.  We thank Alexie Leauthaud and Johannes Lange for a helpful discussion about weak lensing results.

\bibliography{EDE}
\bibliographystyle{mn2e}
\end{document}